\documentstyle{mn}

\input epsf

\begin{document}

\def\Lya{Ly$\alpha\ $}
\def\Lyb{Ly$\beta\ $}
\def\Lyg{Ly$\gamma\ $}
\def\Lyd{Ly$\delta\ $}
\def\Lye{Ly$\epsilon\ $}
\def\LCDM{$\Lambda$CDM\ }
\def\HI{\hbox{H~$\rm \scriptstyle I\ $}}
\def\HII{\hbox{H~$\rm \scriptstyle II\ $}}
\def\DI{\hbox{D~$\rm \scriptstyle I\ $}}
\def\HeI{\hbox{He~$\rm \scriptstyle I\ $}}
\def\HeII{\hbox{He~$\rm \scriptstyle II\ $}}
\def\HeIII{\hbox{He~$\rm \scriptstyle III\ $}}
\def\CIV{\hbox{C~$\rm \scriptstyle IV\ $}}
\def\NHI{N_{\rm HI}}
\def\NHeII{N_{\rm HeII}}
\def\cm2{\,{\rm cm$^{-2}$}\,}
\def\kms{\,{\rm km\,s$^{-1}$}\,}
\def\skm{\,({\rm km\,s$^{-1}$})$^{-1}$\,}
\def\kmsmpc{\,{\rm km\,s$^{-1}$\,Mpc$^{-1}$}\,}
\def\hmpc{\,h^{-1}{\rm \,Mpc}\,}
\def\mpch{\,h{\rm \,Mpc}^{-1}\,}
\def\hkpc{\,h^{-1}{\rm \,kpc}\,}
\def\ev{\,{\rm eV\ }}
\def\kel{\,{\rm K\ }}
\def\intunits{\,{\rm ergs\,s^{-1}\,cm^{-2}\,Hz^{-1}\,sr^{-1}}}
\def\ltsima{$\; \buildrel < \over \sim \;$}
\def\lsim{\lower.5ex\hbox{\ltsima}}
\def\gtsima{$\; \buildrel > \over \sim \;$}
\def\gsim{\lower.5ex\hbox{\gtsima}}
\def\etal{{ et~al.~}}
\def\aj{AJ}
\def\ana{A\&A}
\def\apj{ApJ}
\def\apjs{ApJS}
\def\mn{MNRAS}

\journal{Preprint-01}

\title{The effects of UV background correlations on \Lya forest flux
statistics}

\author[A. Meiksin and M. White]{Avery Meiksin${}^{1}$, Martin White${}^{2}$ \\
${}^1$Institute for Astronomy, University of Edinburgh,
Blackford Hill, Edinburgh\ EH9\ 3HJ, UK \\
${}^2$Departments of Astronomy and Physics, University of California,
Berkeley, CA 94720, USA}


\maketitle

\begin{abstract}
We examine the possible effects of UV background fluctuations on the
pixel flux power spectrum and auto-correlation function of the \Lya
forest due to a finite number of sources in an attenuating medium. We
consider scenarios in which QSO sources dominate the contribution to
the UV background. To estimate their contribution, we use the QSO
luminosity functions from the 2dF and the Sloan Digital Sky Survey. We
estimate self-consistent values for the attenuation length at the
Lyman edge through the Intergalactic Medium using Particle Mesh
simulations of the \Lya forest, normalised by the measured mean \Lya
flux. It is necessary to add the contribution from Lyman Limit Systems
based on their measured statistical properties since the simulations
are unable to reproduce the measured abundances of these systems,
suggesting their formation may involve more complex hydrodynamical
processes than simple collapse into dark matter halos. We examine the
convergence properties of the flux power spectrum and auto-correlation
function under differing assumptions regarding
pressure-smoothing. Convergence to better than 10\% in the flux power
spectrum at proper wavenumbers $k<0.015$\skm requires a comoving
resolution of $30-60\hkpc$ in a comoving box of size $25\hmpc$ at
$z=5$, with the requirements becoming more severe at lower
redshifts. The flux auto-correlation function does not converge to
better than 10\% on scales exceeding 3\% of the box size. We find much
better covergence properties for the {\it relative} effects of the UV
background fluctuations on the flux power spectrum. The background
fluctuations increase the large-scale power and suppress the power at
intermediate scales. At $z\le4$, the effect is only at the few percent
level. By $z>5$, however, the large-scale power is boosted by more
than 50\%, suggesting the flux power spectrum may serve as a useful tool
for distinguishing QSO-dominated UV background scenarios from those in
which more abundant sources like galaxies dominate.
\end{abstract}

\begin{keywords}
methods:\ numerical -- intergalactic medium -- quasars:\ absorption lines
\end{keywords}

\section{Introduction} \label{sec:introduction}

Due in large part to insights gained from hydrodynamical simulations
of structure formation in the universe (Cen \etal 1994; Zhang, Anninos
\& Norman 1995; Hernquist \etal 1996; Zhang \etal 1997; Bond \&
Wadsley 1997; Theuns, Leonard \& Efstathiou 1998) the \Lya forest, as
measured in high redshift Quasi-Stellar Object (QSO) spectra, has
emerged as one of our premier probes of the universe at high
redshift. The results of simulations assuming early homogeneous \HI
and \HeII reionization reproduce the cumulative flux distributions and \HI
column density distributions measured from high resolution spectra to
an accuracy of a few percent (Meiksin, Bryan \& Machacek 2001). More
problematic are the predicted widths of the absorption features, which
appear to require additional sources of broadening, perhaps late \HeII
reionization, to match the measured widths (Theuns \etal 1999;
Bryan \& Machacek 2000; Meiksin \etal 2001).

Parallel to the development of the full hydrodynamical simulations of
the Intergalactic Medium (IGM) are pseudo-hydrodynamical schemes using
pure gravity (Petitjean, M\"ucket \& Kates 1995; Croft \etal 1998;
Gnedin \& Hui 1998; Meiksin \& White 2001). These simpler treatments
are based on two key results from the full hydrodynamical
computations. As a consequence of photoionization equilibrium:\ 1.\ the
gas follows a tight temperature--density relation for the absorbing
material with the neutral hydrogen temperature proportional to the
baryon density to the power of $\sim0.5$ (Zhang \etal 1998; Schaye
\etal 1999), and 2.\ the neutral hydrogen density closely traces the
total matter density on the scales relevant to the forest
($0.1-10\,h^{-1}$Mpc) (Zhang \etal 1998). The structure in QSO
absorption thus tracks, in a calculable way, slight fluctuations in
the matter density of the universe back along the line of sight to the
QSO, with most of the absorption by the \Lya forest arising from gas
with overdensities of a few times the mean density. Although these
techniques are not sufficiently accurate to be used for detailed
modelling of the flux and absorption line properties of the IGM
(Meiksin \& White 2001), they provide a useful means of investigating
convergence questions and the possible role of additional physical
effects on the \Lya forest without the substantial overhead of the
full hydrodynamical treatment. In particular, in recent years they
have been increasingly relied on for predicting the flux power
spectrum of the \Lya forest (Meiksin \& White 2001; Zaldarriaga, Hui
\& Tegmark 2001; Croft \etal 2002b; Seljak, McDonald \& Makarov 2003).

While much work has been done on the physics of the \Lya forest and
the cosmology which may be extracted, the role of fluctuations in the
UV background on the absorption properties of the \Lya forest are
largely unexplored (see Zuo 1992a,b; Fardal \& Shull 1993; Croft et
al. 1999, 2002a; Gnedin \& Hamilton 2002). The effects of
fluctuations in the UV background on the statistical properties of the
\Lya forest were considered previously by Zuo (1992a,b) and Zuo \&
Phinney (1993). Zuo \& Bond (1994) introduced the use of the flux
transmission auto-correlation function for measuring the clustering of
the absorbing gas, although they did not consider the effect of
fluctuations in the UV background apart from a local proximity
effect. As we shall show, the effects of UV background correlations
are most conspicuous only at high redshifts ($z>5$), where the precise
level of absorption by the \Lya forest is most difficult to measure
because of the extremely low flux values, requiring very high
signal-to-noise ratio spectra.

Fardal \& Shull (1993) explored the effects of UV background
fluctuations on the \Lya forest along the lines of Zuo, but using
Monte Carlo realisations of randomly distributed sources and clouds to
estimate the effect of the background fluctuations. Croft et al.~(1999)
estimated the flux power spectrum resulting from UV background fluctuations
generated by randomly distributed QSO sources in a uniform IGM, and
found this to be small, at the few percent level, compared with the
flux power spectrum from the \Lya forest at $z=2.5$. They found that
allowing for QSO clustering had little additional effect. Croft \etal
(2002a) accounted for the UV background fluctuations at $z=3$ assuming
the IGM was ionized by the galaxies in their simulation, and concluded
again the effect on the flux power spectrum is small. Finally Gnedin
\& Hamilton (2002) examined the effect of a fluctuating UV background
at $z=4$ using hydro simulations with pseudo-radiative transfer,
assuming the reionization is accomplished by galaxies in their
simulation. They concluded the fluctuations of the UV background affect
the flux power spectrum at the percent level, but cause an
underestimate of the photo-ionization rate by ${\cal O}(20\%)$. A
comparable boost in the photo-ionization rate is required if QSOs are
the dominant source of the UV background at $z\ge5$ (Paper I).

In this paper, we investigate the effect of the fluctuations on the
2-pt flux distribution (the power spectrum and auto-correlation
function) of the \Lya forest. This extends our earlier work on the
1-pt distribution function reported in Meiksin \& White (2003;
hereafter Paper I). In the next section we describe the simulations
used. In Section 3, we discuss our model for the UV background
fluctuations. The results are presented in Section 4, and we summarise
our conclusions in Section 5. In Appendix A we discuss the
numerical convergence tests we've performed, and we describe our
estimates of the mean \Lya flux in Appendix B.

\section{Numerical simulations} \label{sec:sims}

\begin{table}
\begin{center}
\begin{tabular}{|c|c|c|c|c|c|c|c|} \hline
Model & $\Omega_M$ & $\Omega_v$   & $\Omega_b h^2$ & $h$   & $n$ &
$\sigma_8$ & $\sigma_J$ \\
\hline 
\hline
 1  & 0.30   & 0.70   & 0.020 & 0.70  & 1.05 & 0.97 & 1.59 \\ \hline
 2  & 0.35   & 0.65   & 0.020 & 0.70  & 0.95 & 0.88 & 1.35 \\ \hline
 3  & 0.40   & 0.60   & 0.020 & 0.55  & 1.10 & 0.90 & 1.57 \\ \hline
 4  & 0.30   & 0.70   & 0.022 & 0.70  & 0.95 & 0.92 & 1.32 \\ \hline
 C  & 0.30   & 0.70   & 0.018 & 0.67  & 1.00 & 0.90 & 1.28 \\ \hline
\end{tabular}
\end{center}
\caption{Parameters for the cosmological models.
$\Omega_M$ is the total mass density parameter, $\Omega_v$ the 
vacuum energy density parameter, $\Omega_b$ the baryonic mass fraction,
$h=H_0/100$\kmsmpc, where $H_0$ is the Hubble constant at $z=0$,
$n$ the slope of the primordial density perturbation power spectrum,
$\sigma_8$ the fluctuation normalization at $z=0$ in a sphere of
radius $8h^{-1}$ Mpc, and $\sigma_J$ the fluctuation normalization at $z=3$
filtered on the Jeans scale for gas at $T=2\times10^4\,{\rm K}$.
}
\label{tab:models}
\end{table}

\begin{table}
\begin{center}
\begin{tabular}{|c|c|c|c|c|c|c|c|c|} \hline
$\Omega_M$ & $\Omega_v$   & $\Omega_b h^2$ & $h$   & $n$ &
$\tau$ & $\sigma_8$ & $\sigma_J$ & $\chi^2$ \\
\hline 
\hline
0.30   & 0.70   & 0.020 & 0.70  & 0.95 & 0.10 & 0.912 & 1.31 & 1020.5 \\ \hline
0.30   & 0.70   & 0.020 & 0.70  & 0.95 & 0.15 & 0.958 & 1.38 & 1030.7 \\ \hline
0.30   & 0.70   & 0.020 & 0.70  & 1.00 & 0.10 & 0.954 & 1.46 & 1040.8 \\ \hline
0.30   & 0.70   & 0.020 & 0.70  & 1.00 & 0.15 & 1.000 & 1.54 & 1041.0 \\ \hline
0.30   & 0.70   & 0.022 & 0.70  & 0.95 & 0.15 & 0.920 & 1.32 &  990.1 \\ \hline
0.30   & 0.70   & 0.022 & 0.70  & 1.00 & 0.10 & 0.917 & 1.40 &  992.7 \\ \hline
0.30   & 0.70   & 0.022 & 0.70  & 1.00 & 0.15 & 0.961 & 1.46 &  990.7 \\ \hline
0.30   & 0.70   & 0.024 & 0.70  & 1.00 & 0.10 & 0.881 & 1.33 &  984.0 \\ \hline
0.30   & 0.70   & 0.024 & 0.70  & 1.00 & 0.15 & 0.921 & 1.39 &  979.7 \\ \hline
0.30   & 0.70   & 0.020 & 0.75  & 0.95 & 0.10 & 1.068 & 1.59 & 1212.1 \\ \hline
0.35   & 0.65   & 0.020 & 0.65  & 0.95 & 0.10 & 0.901 & 1.32 & 1001.2 \\ \hline
0.35   & 0.65   & 0.020 & 0.65  & 0.95 & 0.15 & 0.942 & 1.38 & 1008.2 \\ \hline
0.35   & 0.65   & 0.020 & 0.65  & 1.00 & 0.10 & 0.940 & 1.47 & 1042.3 \\ \hline
0.35   & 0.65   & 0.020 & 0.65  & 1.00 & 0.15 & 0.982 & 1.54 & 1039.1 \\ \hline
0.35   & 0.65   & 0.022 & 0.65  & 0.95 & 0.15 & 0.907 & 1.32 &  986.2 \\ \hline
0.35   & 0.65   & 0.022 & 0.65  & 1.00 & 0.10 & 0.903 & 1.40 & 1013.8 \\ \hline
0.35   & 0.65   & 0.022 & 0.65  & 1.00 & 0.15 & 0.946 & 1.47 & 1008.1 \\ \hline
0.35   & 0.65   & 0.024 & 0.65  & 1.00 & 0.10 & 0.866 & 1.34 & 1025.5 \\ \hline
0.35   & 0.65   & 0.024 & 0.65  & 1.00 & 0.15 & 0.907 & 1.40 & 1017.6 \\ \hline
0.35   & 0.65   & 0.024 & 0.70  & 0.95 & 0.10 & 0.987 & 1.49 & 1022.0 \\ \hline
\end{tabular}
\end{center}
\caption{Parameters for cosmological models consistent with WMAP data.
$\Omega_M$ is the total mass density parameter, $\Omega_v$ the vacuum
energy density parameter, $\Omega_b$ the baryonic mass fraction,
$h=H_0/100$\kmsmpc, where $H_0$ is the Hubble constant at $z=0$,
$n$ the slope of the primordial density perturbation power spectrum,
$\tau$ is the Thomson optical depth to the last scattering surface,
$\sigma_8$ the fluctuation normalization at $z=0$ in a sphere of
radius $8h^{-1}$ Mpc, $\sigma_J$ the fluctuation normalization at $z=3$
filtered on the Jeans scale for gas at $T=2\times10^4\,{\rm K}$,
and the values of $\chi^2$ are from fits to the WMAP TT data.
}
\label{tab:WMAP}
\end{table}

To investigate the effects of UV background fluctuations on the power
spectrum of the \Lya forest, we have performed simulations of 5
different cosmological models within the $\Lambda$CDM family. In each
case we used pure Particle Mesh (PM) dark matter simulations,
mimicking the temperature of the gas using a polytropic equation of
state and assuming gas and dark matter have the same spatial
distribution.  This has been shown (Petitjean, M\"ucket \& Kates 1995;
Croft \etal 1998; Gnedin \& Hui 1998; Meiksin \& White 2001), to
produce results comparable to the full hydrodynamical simulations at
the level of 10 per cent. The agreement is expected to be even better
on the large scales of interest to this work. The parameters for the
simulations are provided in Table~\ref{tab:models}. The models are
all for a flat universe and are consistent with \Lya forest
constraints on matter fluctuations $\sigma_J$ on the Jeans scale (Meiksin,
Bryan \& Machacek 2001). Models 1--3 were run previously to the WMAP
results and were constructed to be consistent with existing
large-scale structure, cluster abundance, and CMB constraints at the
time. Model 4 was later added explicitly to be consistent with WMAP
results as well (Verde et al.~2003, Hinshaw et al.~2003, Kogut et
al.~2003). Although Models 1--3 are not the best matches to the recent
WMAP data, they are close to models that are good matches on the
scales relevant to the \Lya forest. We provide in Table \ref{tab:WMAP}
a list of similar models with their $\chi^2$ agreement to the WMAP TT
data. Model C is the concordance model of Ostriker \& Steinhart
(1995). A description of the parallel PM code used is given in Paper
I. In each case we used $512^3$ particles and a $1024^3$ force mesh,
in a cubic box with (comoving) side length $25\,h^{-1}$Mpc, except for
Model C, for which the comoving box size was
$30\,h^{-1}$Mpc. Convergence tests, described in Appendix A, suggest
that these scales are adequate for our purposes.

\begin{table}
\begin{center}
\begin{tabular}{|c|c|c|c|c|c|} \hline
$z$ & $\langle\exp(-\tau)\rangle$ & $\Gamma_{-12}$ & $r_0$ & $N_0$ & $N_{\rm box}$ \\ \hline\hline
6.0 &   $<0.006$  &  $<0.14$ & $<23.0$ & 3.9  & 1.2 \\
5.5 &   $0.079^{+0.017}_{-0.013}$  & $0.37^{+0.06}_{-0.05}$ & $77^{+13}_{-9}$ & 130  & 1.0 \\
5.0 &   $0.12^{+0.03}_{-0.04}$  & $0.31^{+0.07}_{-0.09}$ & $85^{+17}_{-22}$ & 200   & 1.2 \\
4.0 &   $0.36\pm0.03$  & $0.43^{+0.06}_{-0.05}$ & $80\pm5$  & 240 & 1.7 \\
4.0 &   $0.47\pm0.03$  & $0.76^{+0.12}_{-0.11}$ & $104\pm6$  & 510 & 1.7 \\
3.89 &  $0.48\pm0.02$  & $0.68^{+0.08}_{-0.07}$ & $106\pm5$  & 560 & 1.8 \\
3.0 &   $0.70\pm0.02$  & $0.88^{+0.14}_{-0.12}$ & $211\pm11$  & 5900 & 2.4 \\
2.75 &  $0.74\pm0.04$  & $0.86^{+0.36}_{-0.24}$ & $250\pm28$  & $1.1\times10^4$ & 2.6 \\
\end{tabular}
\end{center}
\caption{Some of the output times for the simulations. For each redshift the
mean \Lya flux, $\langle\exp(-\tau)\rangle$, the required \HI
photoionization rate, in units of $10^{-12}\,{\rm s^{-1}}$, to recover the
mean flux assuming a homogeneous UV background, the attenuation length $r_0$,
in comoving $h^{-1}$Mpc, and the mean number of sources per attenuation
volume, $N_0$, are listed for Model 4. The last
column, $N_{\rm box}$, shows the number of sources in the simulation volume of
side $25\,h^{-1}$Mpc (comoving). The two values at $z=4.0$ correspond to two
disparate estimates of the mean \Lya flux.}
\label{tab:aout}
\end{table}

Given a set of final particle positions and velocities, we compute the spectra
as follows.  First the density and density-weighted line-of-sight velocity
are computed on a grid (using CIC interpolation) and smoothed using FFT
techniques. We try three different smoothing schemes, as described below and
in Appendix \ref{sec:tests}. The default scheme simply smooths with a Gaussian
of width one grid cell.
This forms the fundamental data set.
A regular grid of $32\times 32$ sightlines is drawn through the box,
parallel to the box sides. Along each sightline we integrate (in real space)
to find $\tau(u)$ at a given velocity $u$.  Specifically we define
\begin{equation}
  \tau(u) = \int dx\, A(x) \left[ {\rho(x)\over\bar{\rho}} \right]^2
            T(x)^{-0.7} b^{-1} e^{-(u-u_0)^2/b^2}
\label{eqn:taudef}
\end{equation}
where $u_0=xaHL_{\rm box}+v_{\rm los}$ and $b=\sqrt{2k_B T/ m_{\rm H}}$ is the
Doppler parameter, where $m_{\rm H}$ is the mass of a hydrogen atom.
The flux at velocity $u$ is $\exp(-\tau)$. The integration variable $x$
indicates the distance along the box in terms of the expansion velocity across
the box.
In evaluating Eq.~(\ref{eqn:taudef}) we assume a power-law relation between
the gas density and temperature
\begin{equation}
  T \equiv T_0 \left( {\rho\over\bar{\rho}} \right)^{\gamma-1}
\end{equation}
In practice, $T_0$ and $\gamma$ will depend on the reionization
history of the IGM, but we shall fix $T_0=2\times 10^4$K and $\gamma=1.5$
throughout as described in Paper I.

In terms of the baryon density parameter $\Omega_b$, hydrogen baryonic mass
fraction $X$, the metagalactic photoionization rate $\Gamma_{-12}$, in units
of $10^{-12}\, {\rm s^{-1}}$, and the number $f_e$ of electrons per
hydrogen nucleus, the function $A(x)$ is given by:
\begin{equation}
  A(x)=18.0f_e\left(\frac{X}{0.76}\right)\left(\frac{\Omega_bh^2}{0.02}\right)
  \Gamma_{-12}^{-1}(x)L_{\rm box}(1+z)^5,
\label{eq:Adef}
\end{equation}
where $\Gamma_{-12}(x)$ is evaluated at position $x$, $L_{\rm box}$ is in Mpc
and $b$ in eq.~(\ref{eqn:taudef}) is in \kms.
We normalise the spectra according to measurements of the mean absorbed
flux $\langle\exp(-\tau)\rangle$ as reported in Fan \etal (2002)
(their Figure 1) for $z\ge4$. We discuss the values adopted at other
redshifts in Appendix B.
Values for $\langle \exp(-\tau)\rangle$ at each of the output times of the
simulation are provided in Table \ref{tab:aout}.

In order to compute the flux power spectra, it is necessary to filter
the simulation data to reduce the effects of particle noise at the
grid level. The minimal amount of filtering we consider is Gaussian
smoothing on the scale of a mesh cell. While this is appropriate for
computing the power spectrum of the dark matter, it neglects any
`natural' filtering arising from pressure forces acting on the
baryons. The most accurate means of predicting the power spectrum is
by performing full hydrodynamical simulations at both high spatial
resolution and in a box of sufficient size to ensure convergence. Such
calculations, however, are still prohibitively expensive to repeat for
a range of models, and also because the origin of the line broadening
is not fully understood (Theuns \etal 1999; Bryan \& Machacek 2000;
Meiksin \etal 2001), requiring additional modelling which affects both
the shape and amplitude of the predicted flux power spectrum. As we
are primarily interested in the {\it relative} effect of fluctuations
in the UV background, we do not consider an exhaustive range of
smoothing procedures, but only consider a few straightforward ones to
judge how important smoothing is to our final results. Following
Gnedin \& Hui (1998), in addition to the minimal smoothing models, we
create models with additional smoothing on scales of the order of the
Jeans length. Specifically we compute a `Jeans scale',
$k_J=(3/2)^{1/2}c_s^{-1}H(z)/(1+z)$, where $c_s$ is the sound speed,
using the assumed temperature and equation of state of the gas and
used this in place of our one mesh cell filtering above. We consider
two different types of smoothing, a Gaussian of width
$\sigma=2^{1/2}k_J$ and a form motivated by linear perturbation
theory, viz.~$(1+[k/k_J]^2)^{-1}$. We emphasize again that all of
these approaches, like the pseudo-hydrodynamical approach employed by
McDonald (2003), are ad hoc and can only be justified by comparison to
true hydrodynamic simulations, but modify the power mostly on scales
where thermal broadening is also very important.

\section{UV background}

\subsection{UV intergalactic attenuation}

For the uniform background case, normalising the simulation results by
the mean \Lya flux also fixes the attenuation properties of the IGM
since the ionization structure of the IGM has been fixed by the
normalisation. This was exploited in Paper I to place constraints on
the attenuation length $r_0$ and mean emissivity required to recover
the measured values of the mean \Lya flux for $z>4$. At lower
redshifts, the simulation results are no longer sufficient for
determining the Lyman limit optical depth of the IGM. We find that the
simulations fail to reproduce the measured number of Lyman Limit Systems
(LLSs) by a factor of several. A similar conclusion was reached by
Gardner \etal (1997) on the basis of lower resolution simulations than
those we present. There, the authors attempted to correct for the
missing LLSs by estimating the contribution from unresolved halos on
the basis of Press-Schechter theory. They found that they still fell
well short of the observed number. Our simulations resolve halos down
to masses of about $10^8-10^9\,M_\odot$, corresponding to circular
velocities smaller than $20-25$\kms at $z=3$. The baryons in smaller
mass halos would be heated to temperatures too high to remain bound
after photoionization (Meiksin 1994), so that smaller mass halos would
seem unlikely candidates for LLSs. We again find too few LLSs. This
suggests that the origin of these systems may involve more complex
hydrodynamical processes than simple collapse into dark-matter halos,
such as Jeans fragmentation in the halos (Meiksin 1994), tidal streaming
or galactic winds.

In order to estimate the attenuation length, we adopt the measured abundance
of LLSs from Stengler-Larrea \etal (1995) for $\tau_L>1$,
$dN/dz=N_0(1+z)^\gamma$, with $N_0=0.25^{-0.10}_{+0.17}$ and
$\gamma=1.50\pm0.39$ for $0.3\lsim z\lsim4$. The contribution of the
LLSs to the effective optical depth of the IGM at the Lyman edge is given by
(Zuo 1992b)
\begin{equation}
\tau_{\rm eff}^L=\int_{z_{\rm obs}}^{z_{\rm em}}dz \int_1^\infty d\tau_L
\frac{\partial^2N}{\partial z \partial\tau_L}\left\{1-\exp\left[-\tau_L
\left(\frac{1+z_{\rm obs}}{1+z}\right)^3\right]\right\}.
\label{eq:tauL}
\end{equation}
Here, the optical depth corresponds to the path between the source at
$z_{\rm em}$ and the position of observation at $z_{\rm obs}$.
Because of the difficulty in measuring the
high optical depths involved, the actual distribution of individual
Lyman limit optical depths is less certain. Adopting a power law distribution
$\partial^2/\partial z\partial\tau_L=A(1+z)^\gamma\tau_L^{-\beta}$, and
assuming a flat universe so that $dz/dr_p\approx(H_0\Omega_M^{1/2}/c)
(1+z)^{5/2}$ where $r_p$ is proper length, we obtain
\begin{equation}
\frac{d\tau_{\rm eff}^L}{dr_p}\approx\frac{H_0\Omega_M^{1/2}}{c}N_0
(1+z)^{5/2+\gamma}\left[1-e^{-1}+\Gamma(2-\beta,1)\right],
\end{equation}
where $\Gamma(a,x)$ is the incomplete $\Gamma-$function.
Adopting $\Omega_M=0.3$, $h=0.7$, $N_0=0.25$, $\gamma=1.50$, we obtain
$d\tau_{\rm eff}^L/dr_p\approx3.2\times10^{-5}{\cal T}(1+z)^4$,
where ${\cal T}=0.85$ for $\beta=2$ and 0.91 for $\beta=1.5$. In the limit of
an infinitely steep distribution, so that all systems have $\tau_L=1$,
${\cal T}=0.63$.

To estimate the total Lyman limit optical depth through the IGM, we
add the LLS contribution to the values we find from a given simulation
for $z\le4$.  Current measurements do not permit an estimate of the
LLS contribution at higher redshifts. Given the uncertain origin of
these systems, we conservatively do not include them in our estimates
for $z>4$. Doing so would result in a decrease of the attenuation
length. We also assume the proper source emissivity does not evolve in
our estimate of the attenuation length. Over the range of measured
optical depths, this assumption only weakly affects our results.

\begin{figure}
\begin{center}
\leavevmode \epsfxsize=3.3in \epsfbox{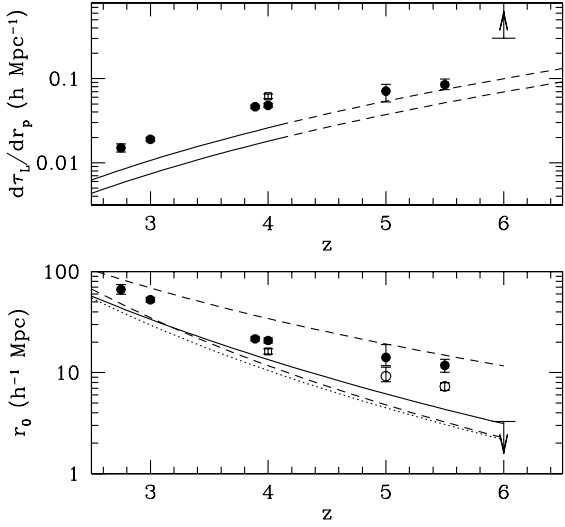}
\end{center}
\caption{The (proper) gradient $d\tau_L/dr_p$ of the Lyman limit optical depth
(upper panel) and the (proper) attenuation length (lower panel) for Model 4.
The points are estimated from the simulation results
using the observational constraints on the mean \Lya flux.
The contribution from Lyman Limit Systems is included in the values of the
attenuation length for $z\le4$. The effect Lyman Limit Systems would have
on the attenuation length at $z=5$ and 5.5 is shown by the open points.
The open square at $z=4$ corresponds to an
alternative lower estimate for the average absorbed flux (see text).
The two solid curves in the upper panel show the upper and
lower limits for the measured contribution from Lyman Limit Systems, with the
dashed extensions extrapolations based on the lower redshift measurements.
The two dashed lines in the lower panel are based on the low and
high attenuation models of Meiksin \& Madau (1993), and the solid line
is based on their medium attenuation model. The dotted line
is based on the attenuation estimate of Haardt \& Madau (1996).
}
\label{fig:r0}
\end{figure}

In Fig.\ref{fig:r0}, we show the redshift evolution of $d\tau_{\rm
eff}^L/dr_p$ from the simulation results for Model 4, along with the
range in the LLS contribution based on observations. We also show the
evolution of the attenuation length, defined by $r_0=(d\tau_{\rm
eff}^L/dr_p)^{-1}$, and compare our estimate with previous estimates
based on direct \Lya forest line counts from Meiksin \& Madau (1993)
and Haardt \& Madau (1996). We have included the LLS contribution to
the attenuation lengths for $z\le4$. We also show the effect including
the LLS contribution, if extrapolated to $z\ge5$, would have on $r_0$
by the open points at $z=5$ and 5.5. Our estimates for the attenuation
length lie between the medium and low attenuation models of Meiksin \&
Madau (1993), and lie systematically above the estimate of Haardt \&
Madau (1996). The redshift dependence of the (proper) attenuation
lengths we find is fit by $r_0\approx6800\hmpc (1+z)^{-3.5}$ over
$2.75\le z\le5.5$ to $\sim20\%$ accuracy.  The attenuation lengths
allowing for an extrapolation of the LLS contribution to $z>4$ are fit
over the same redshift range by $r_0\approx1.7\times10^4\hmpc
(1+z)^{-4.2}$ to $\sim10\%$ accuracy. We caution against extending the
fits outside the specified redshift range. In particular, we find the
attenuation length decreases precipitously between $z=5.5$ and 6, a
consequence of the observed sharp rise in the \Lya optical depth at
these redshifts (Fan \etal 2002).

\subsection{Source contribution}

\begin{figure}
\begin{center}
\leavevmode \epsfxsize=3.3in \epsfbox{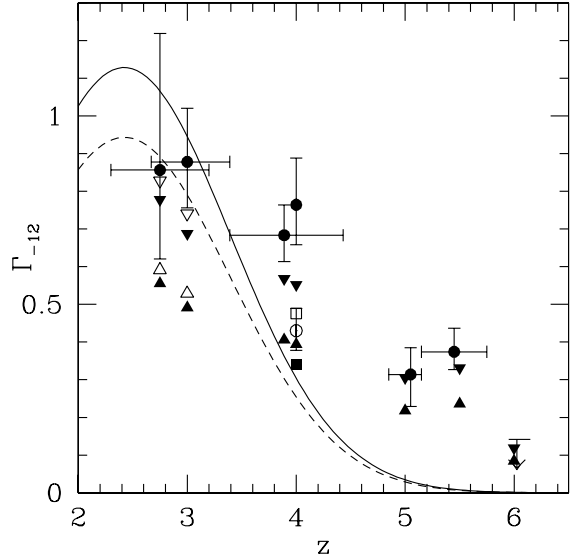}
\end{center}
\caption{A comparison between the estimated \HI metagalactic
photoionization rate $\Gamma_{-12}$ from QSO sources and the rate
required to match the measured mean \Lya flux, as estimated for Model
4. The filled data points with error bars are from Table
\ref{tab:aout}. The open data point at $z=4$ is for an alternative
lower estimate of the mean \Lya flux (see text).  The estimated rates
for QSO sources, shown by the filled triangles, assume a QSO
luminosity function with a steep end slope of $\beta_1=3.2$. The
inverted filled triangles include a 40\% boost, nearly the maximum possible
contribution from diffuse recombination radiation emitted by the
IGM. The open triangles and inverted triangles are analogous estimates
assuming $\beta_1=3.41$. The filled square at $z=4$ is for $\beta_1=3.2$
and the alternative lower mean \Lya flux considered. The open square
includes a 40\% boost in this estimate due to diffuse radiation from
the IGM. Also shown are the estimates from Haardt \& Madau (1996)
based on earlier QSO source counts and assuming an intrinsic QSO spectral index
shortward of the Lyman edge of $\alpha_Q=1.5$ (solid line)
and $\alpha_Q=1.8$ (dashed line), and
which include the diffuse recombination radiation from the IGM.
}
\label{fig:G12}
\end{figure}

\begin{figure}
\begin{center}
\leavevmode \epsfxsize=3.3in \epsfbox{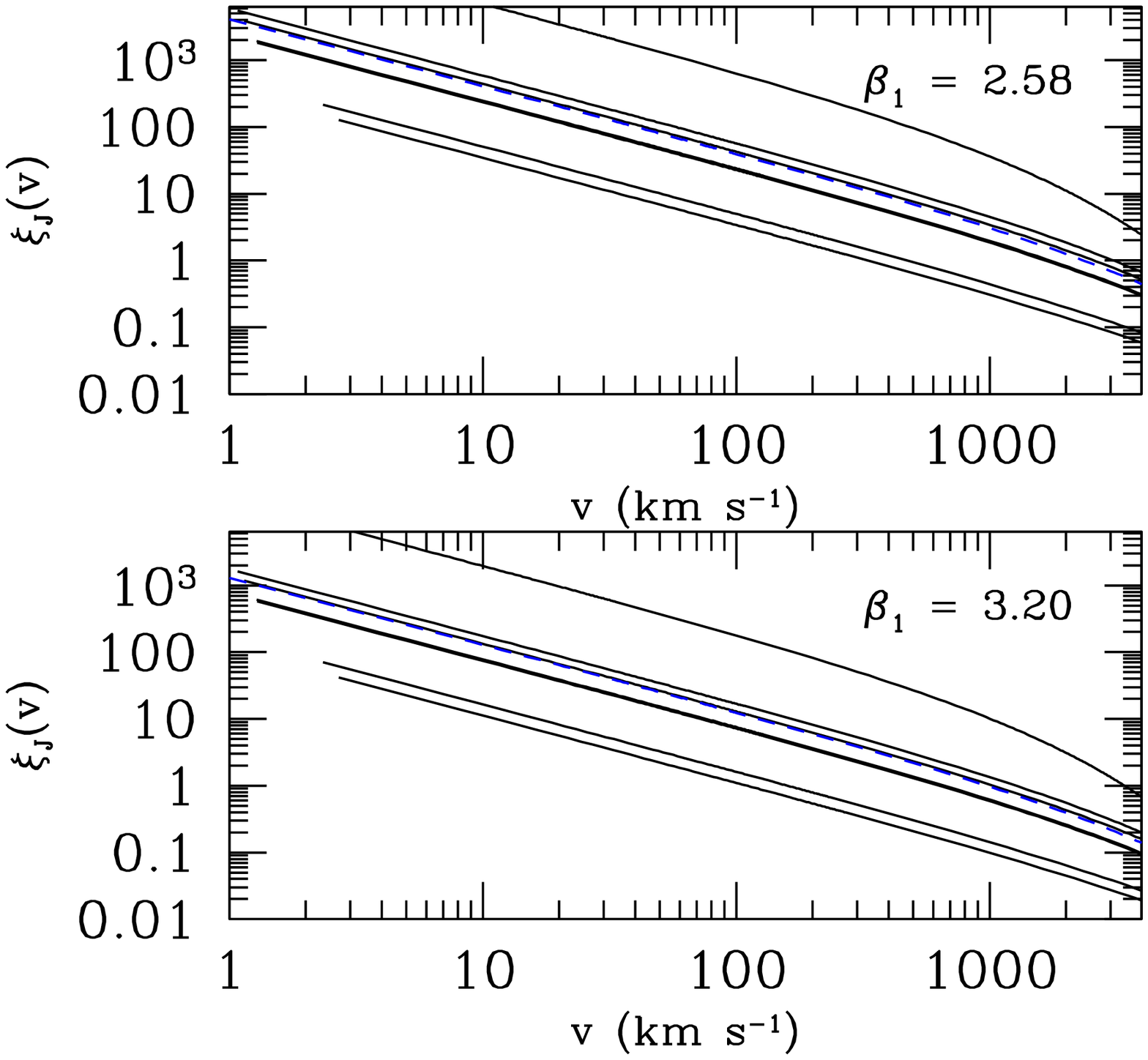}
\end{center}
\caption{The auto-correlation function in the UV background predicted
from QSO sources with $\beta_1=2.58$ (upper panel) and $\beta=3.20$
(lower panel). The solid curves, from top to bottom, correspond to the
redshifts $z=6$, 5.5, 5, 4, 3.89, 3, and 2.75. The $z=4$ solid curve is
computed for a (comoving) attenuation length of $r_0=104 h^{-1}\, {\rm Mpc}$
and nearly coincides with the $z=3.89$ result. Also shown is the result at
$z=4$ assuming $r_0=80 h^{-1}\, {\rm Mpc}$ (dashed line), which nearly
coincides with the $z=5$ result.}
\label{fig:xiJ}
\end{figure}

The QSO luminosity function has a slope at the bright end varying as
$\phi(L)\propto L^{-\beta_1}$, where $\beta_1=3.41$ according to the
2dF QSO sample for $z<2.3$ (Boyle \etal 2000), and $\beta_1=2.58$
according to the Sloan QSO sample for $z>3.5$ (Fan \etal 2001), although
$\beta_1=3.2$ is still permitted by the data.
We use the procedure in Paper I for adjusting
the parameters of the Boyle, Shanks \& Peterson (1988) luminosity function,
\begin{equation}
\Phi(L)=\frac{\Phi^*}{\left[(L/L^*)^{\beta_1}+(L/L^*)^{\beta_2}\right]},
\label{eq:bqlf}
\end{equation}
viz., by fixing $\beta_2=1.58$, in accordance with Boyle \etal (2000),
and matching the SDSS counts for $M_{1450}<-25.7$ for $z<6$ and for
$M_{1450}<-27.1$ at $z=6$ (and allowing for differences in the assumed
values of the Hubble constant). We show the predicted contribution of
QSOs to $\Gamma_{-12}$ in Fig.~\ref{fig:G12} for $\beta_1=3.2$, and
compare these with the required rates from Table \ref{tab:aout}. We
have followed the procedure of Paper I for estimating $\Gamma_{-12}$
from the QSO emissivity, including hardening of the metagalactic
radiation field on filtering through the IGM (Haardt \& Madau
1996). Specifically, for $z\le3$ we assume a metagalactic spectral
index $\alpha_{\rm MG}=0$, while for $3.89\le z\le5.5$ we take
$\alpha_{\rm MG}=-1.3$, except at $z=5.0$ ($\alpha_{\rm MG}=-0.5$) and
$z=6$ ($\alpha_{\rm MG}=0$). These values were chosen to recover the
required photoionization rates within $2\sigma$, though usually the
addition of some diffuse radiation is required.  The value of
$\alpha_{\rm MG}$ is uncertain and can only be computed fully
self-consistently by solving the detailed radiative transfer problem
through the high redshift IGM, but we do not expect it to deviate much
from the values we use. In any case, the spectral index only weakly
affects the overall ionization rate after integrating over the
photoelectric cross-section. Significant diffuse radiation resulting
from re-radiation by the IGM through radiative recombinations is
expected at these high redshifts (Meiksin \& Madau 1993; Haardt \&
Madau 1996), with as much as a $(\alpha_A-\alpha_B)/\alpha_A\approx
40-50\%$ boost in the total metagalactic
photoionization rate over the direct contribution from the QSO
sources. We show the enhanced rates for a 40\% boost in Fig.~\ref{fig:G12}.

The QSO contribution to the metagalactic photoionization rate lies
systematically below the required values at $z=3-4$. Given the
uncertainties in the required rates, as well as in the QSO counts at
these redshifts, it is unclear how seriously the discrepancy should be
taken. Based on a different estimate of the QSO counts, Haardt \&
Madau (1996) compute a photoionization rate that well matches the
required rate at $z\approx3$ (see Fig.~\ref{fig:G12}), although their
value lies much below the required rate by $z=4$. (A required slower
rate of decline at $z\gsim3$ than predicted by Haardt \& Madau was
also found by Meiksin, Bryan \& Machacek 2001.) A boost in
the IGM temperature at these redshifts, as expected for late \HeII
reionization (Madau \& Meiksin 1994; Reimers \etal 1997; Meiksin,
Bryan \& Machacek 2001), would also help ease the discrepancy (because
of the resulting reduced hydrogen recombination rate). Finally, if the
luminosity from QSO sources is beamed, the estimated UV contribution
would increase like the ratio of the UV beam solid angle to the
optical beam solid angle, since the QSO counts are based on restframe
optical or near UV (longward of the Lyman edge) surveys. (In
principle, this could also result in a decrease in the estimated QSO
contribution to the UV background if the ratio were less than unity.)
None the less, it is interesting
in this context that Steidel, Pettini \& Adelberger (2001) report a
large Lyman-continuum emissivity from Lyman-break galaxies at
$z\simeq3.4$. Adjusting their value down by the factor
$\Omega_M^{1/2}$ for Model 4, the predicted photoionization rate at
$z=3$, assuming a non-evolving proper emissivity, is
$\Gamma_{-12}=4.4$, well in excess of the required value in Table
\ref{tab:aout}. Within the context of our models, such a high rate
would so reduce the neutral fraction of the \HI that the IGM would
become significantly more transparent to \Lya photons than
measured. To match our estimated $3\sigma$ upper limit on the
photoionization rate would require the Lyman-continuum emitting phase
of the Lyman-break galaxy population to last only over a redshift
interval $\Delta z\approx0.1$. Thus the required excess
photoionization above the QSO contribution could be accounted for by
the Lyman-break galaxies, but only if they are a bursting population
at these redshifts. Based on population synthesis models of the
galaxies, Ferguson, Dickinson \& Papovich (2002) argue this indeed may
be the case.

\subsection{UV background fluctuations}

We compute the auto-correlation function of the QSO contribution
to the UV background, using the formalism of Zuo (1992b) (also see
Paper I), both for $\beta_1=2.58$ and 3.2. The results are shown in
Fig.~\ref{fig:xiJ}. Formally the correlation function diverges at
zero separation, since the variance in the radiation field diverges
(Zuo 1992a; Paper I). The typical width of an absorption feature is
25\kms, so when the correlations become small on this scale, we expect
them to have a negligible influence on the flux auto-correlations or
power spectra. We find that for $z>3$, the correlations still exceed
unity on this scale, so that we may expect the effect of UV background
fluctuations to imprint a signature on the flux statistics of the \Lya
forest. We note that for $\beta_1=2.58$, additional sources (like
galaxies) must be introduced to recover the required \HI
photoionization rate (Paper I), but that these sources would only
dilute the correlations by a factor of 2--3.

\begin{figure}
\begin{center}
\leavevmode \epsfxsize=3.3in \epsfbox{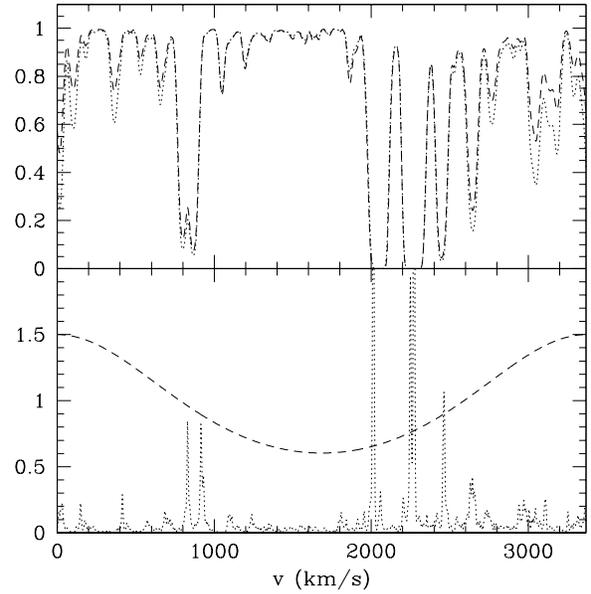}
\end{center}
\caption{A simulated spectrum from a `typical' sightline at z=5 from Model 1.
The upper panel shows the spectrum with a uniform ionizing background
($A(x)\equiv $const, dashed) and where the UV background comes from a single
source situated at the center of the box (dotted).
The lower panel shows $0.1\times\rho/\bar{\rho}$ (dotted) and $A(x)/\bar{A}$
(dashed) for the single source case.}
\label{fig:spectrum}
\end{figure}

The effect of a discrete source on the \Lya flux is shown in
Fig.~\ref{fig:spectrum}.  Here we artificially suppose that the entire
UV background flux comes from a single source located in the center of
the simulation box. For a line-of-sight which passes slightly off
center the $A(x)$ and spectrum are as shown, along with the spectrum
that would have resulted along this sightline if the UV background
were completely uniform.

To model the fluctuations in the UV background from a `realistic'
population of discrete sources we resort to a Monte Carlo
procedure. We create random source lists, with luminosities drawn from
the QSO luminosity function above.  A random number of QSOs is chosen
consistent with the luminosity density of QSOs with absolute magnitude
at restframe 1450A $M_{1450}<-18.5$. The luminosity of each QSO is
drawn at random from the luminosity function, and the sources are
distributed at random positions in a volume larger than the size of
the simulation box. We choose the box size to be sufficiently large to
include the full effects of attenuation but not so large that redshift
or evolution effects become important, so that the box is not larger
than the effective size over which sources contribute to the local UV
intensity (no larger than $l_U$ in Paper I).

The frequency specific flux at the Lyman edge is then calculated at each
position on the simulation grid from
\begin{equation}
  f_L = \sum_i L_i {e^{-r_i/r_0}\over {4\pi r_i^2}}
\end{equation}
where the sum is over the sources, the specific luminosity of source
$i$ at the Lyman edge is $L_i$, and $r_i$ is the distance of the $i$th
source from the fiducial point in the simulation volume. To convert
from the QSO luminosity at 1450A to the Lyman edge, we adopt a QSO
spectral shape of the form $f_\nu\sim\nu^{-0.99}$ from $1050-1450$A
and $f_\nu\sim\nu^{-1.8}$ from $912-1050$A (Zheng \etal 1997).  Since
our simulation box, and the relevant attenuation lengths, is small we
do not keep track of source lifetimes or lightcone effects (see Croft
2003).  {}To compute the total rate $\Gamma_{-12}$ of ionizing photons
(in units of $10^{-12}\, {\rm s^{-1}}$), from $f_L(x)$, we adjust the
spectral index of the metagalactic background to reproduce the target
values of $\langle\exp(-\tau)\rangle$ in Table~\ref{tab:aout}. As in
the uniform UV background case, for $z\ge3.89$ the required values are
generally negative ($-1.3\lsim\alpha_{\rm MG}\lsim-0.5$), consistent
with the expected hardening of the radiation field near the Lyman edge
on being filtered through the IGM.  Other sources, like galaxies, may
of course contribute as well. Adjusting $\alpha_{\rm MG}$ to reproduce
the required ionization rate provides an upper limit to the possible
effects of UV background correlations resulting from attenuation; the
introduction of additional sources would tend to dilute these
effects. We also consider cases with a diffuse radiation field, as is
expected to arise from radiative recombinations within the IGM.

Once we have $\Gamma_{-12}$ throughout the simulation volume, we
compute $A(x)$. We ignore the fluctuations in the IGM temperature that
would result, noting that this is another possible contribution to the
flux power spectrum, but one again that requires detailed modelling of
the reionization process.  Since our simulation volume is small, we
explicitly checked that imposing periodic boundary conditions on the
flux distribution modified our results by much less than 1 per cent.

We run sufficient realizations of this procedure to ensure convergence
(one to several hundred, depending on redshift), keeping track of the
UV background and flux correlation functions and the flux power
spectrum. This enables us to compute both the mean spectra and their
correlations.

\section{Results}

\subsection{Pixel flux distribution}

\begin{figure}
\begin{center}
\leavevmode \epsfxsize=3.3in \epsfbox{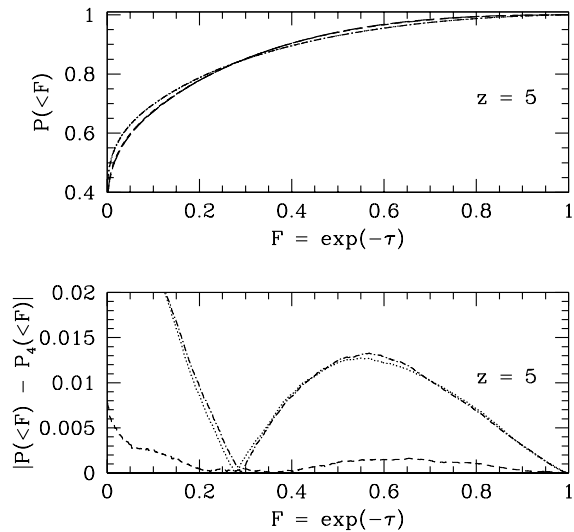}
\end{center}
\caption{(Top panel) The cumulative distributions of the pixel flux
at $z=5$ for the four models assuming no UV background fluctuations.
The models are labelled as Model 1 (dotted), Model 2 (short dashed),
Model 3 (dot-short dashed), Model 4 (long dashed). The distributions
group according to the value of $\sigma_J$, flattening as $\sigma_J$
increases. (Bottom panel) The
differences between the cumulative flux distributions for
Models 1--3 and the distribution for Model 4 assuming no UV background
fluctuations at $z=5$. (The line types correspond to those in the
top panel.)}
\label{fig:cumd}
\end{figure}

\begin{figure}
\begin{center}
\leavevmode \epsfxsize=3.3in \epsfbox{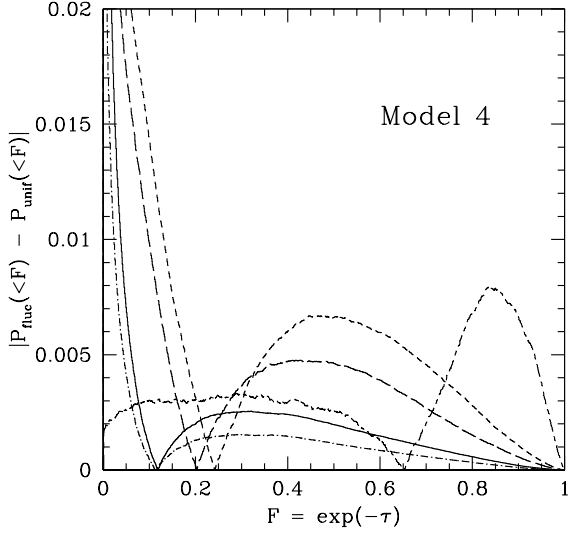}
\end{center}
\caption{The differences between the cumulative
flux distributions for Model 4 with and without UV background fluctuations,
at $z=4$ (short-long dashed), 5 (short dashed), 5.5 (long dashed) and
6 (solid). Also shown is the difference at $z=6$ allowing for a diffuse
component added to the fluctuating UV background (dot-short dashed).}
\label{fig:cumdfluc}
\end{figure}

The cumulative pixel flux distributions at $z=5$ are shown in
Fig.~\ref{fig:cumd}. The slopes of
the distributions are determined essentially by the value of $\sigma_J$,
and show a systematic trend of flattening with increasing
$\sigma_J$. The same trend was found by Meiksin \etal (2001) at $z=3$,
who exploited it to place stringent constraints on $\sigma_J$ from high
spectral resolution measurements of the \Lya forest. These results corroborate
the trend found there. For a single high resolution spectrum of typically
$N_{\rm pix}=10^4$ pixels, it is possible to distinguish the cumulative
distributions only to differences of about $1/N_{\rm pix}^{1/2}=0.01$, or
somewhat larger allowing for pixel flux correlations (Meiksin, Bryan \&
Machacek 2001), so that while Models 1 and 3 could be distinguished from
Model 4, Model 2 could not on the basis of a single spectrum.

The effect of fluctuations in the UV background on the cumulative flux
distributions for Model 4 at $z=4$, 5, 5.5 and 6 is shown in
Fig.~\ref{fig:cumdfluc}. As found in Paper I, the UV background fluctuations
produce only a small distortion on the pixel flux distribution. The distortions
found here are somewhat larger than found in Paper I, because here we now also
account for correlations in the UV background. As found in Paper I, the effects
of the UV background fluctuations are generally smaller than or comparable to
the uncertainty between models with differing values of $\sigma_J$. Only
at the smallest flux values ($<0.1$) would it be possible to distinguish the
effects of the fluctuations from an uncertainty in the model if the model is
known with sufficient accuracy. A comparable level of uncertainty is introduced
by the uncertain equation of state of the gas (Paper I), but this may be
removed by adjusting the temperature to match the measured distributions of
Doppler parameters or wavelet coefficients (Meiksin 2000;
Meiksin, Bryan \& Machacek 2001).

As in Paper I, we find that the demands on the mean photoionization
rate increase in the presence of UV background fluctuations. Allowing
for the correlations in the UV background, however, reduces the
effect, and it actually inverts by $z=6$. Specifically, for $z\le5.5$
we find the mean photoionization rate must increase by 5--10\% when
the (correlated) UV background fluctuations are included in the
models. The amount of the increase depends on the fractional
contribution arising from a diffuse radiation field and on the
redshift. At $z=6$, the required mean photoionization rate is instead
reduced by 10-20\% when the now strongly correlated UV background
fluctuations are included.

\subsection{Pixel flux power spectrum}

\begin{figure*}
\epsfbox{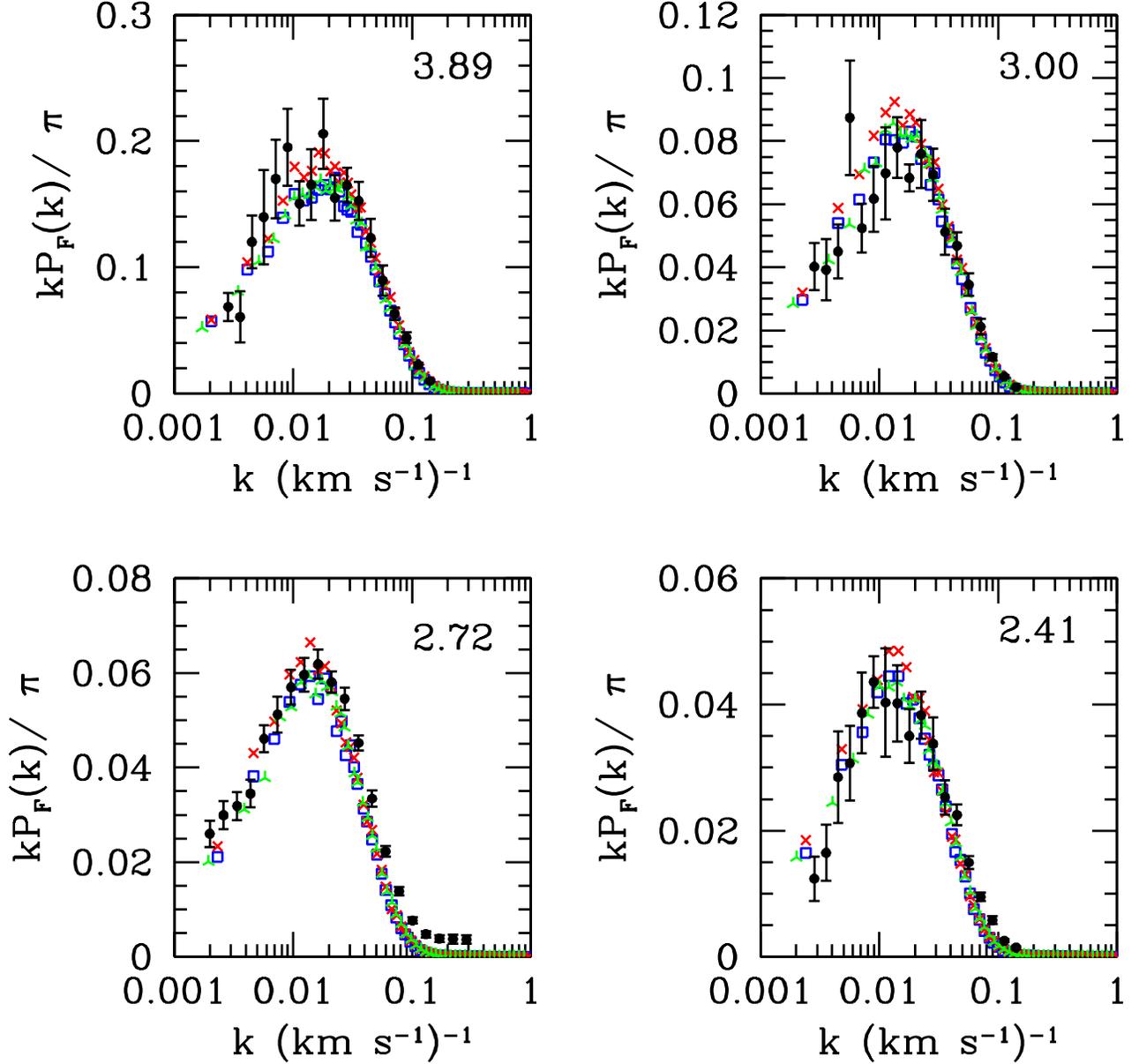}
\caption{The flux power spectrum $P_F(k)$, as a function of
restframe velocity wavenumber, for Models 1 ($\times$) and 4 (square) and
the concordance model (inverted Y), compared with the measurements in high
resolution spectra at the indicated redshifts.
}
\label{fig:PFk_data}
\end{figure*}

\begin{figure}
\begin{center}
\leavevmode \epsfxsize=3.3in \epsfbox{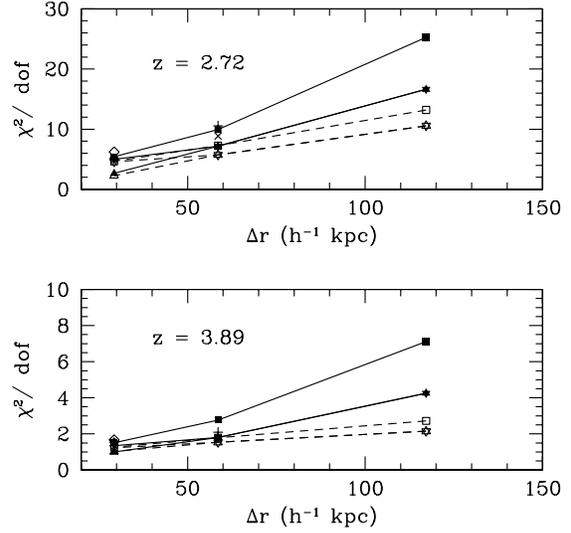}
\end{center}
\caption{The $\chi^2$ per degree of freedom between the concordance
model and the measured flux power spectrum $P_F(k)$, as a function
of (comoving) resolution. The upper panel is for z=2.7, and
the lower for z=3.9. The open symbols use all points with
$k<0.015$\skm, and the filled symbols use all points with $k<0.032$\skm.
The results are shown for the $30\, {\rm h^{-1}\, Mpc}$ (comoving) box
simulations, for Gaussian smoothing on the mesh scale (triangle),
Gaussian Jeans smoothing (inverted triangle), and Gaussian linear smoothing
(square). Also shown is the Gaussian mesh-size smoothing case for the
$60\, {\rm h^{-1}\, Mpc}$ (comoving) box simulation for $k<0.015$\skm
($+$) and $k<0.032$\skm ($\times$), as well as the results using alternative
equations of state (open and filled diamonds) for the
$30\, {\rm h^{-1}\, Mpc}$ high resolution box (see text).
To help guide the eye, filled symbols are
connected by solid lines and open symbols by dashed lines.
}
\label{fig:PFk_data_chi2}
\end{figure}

\begin{figure*}
\epsfbox{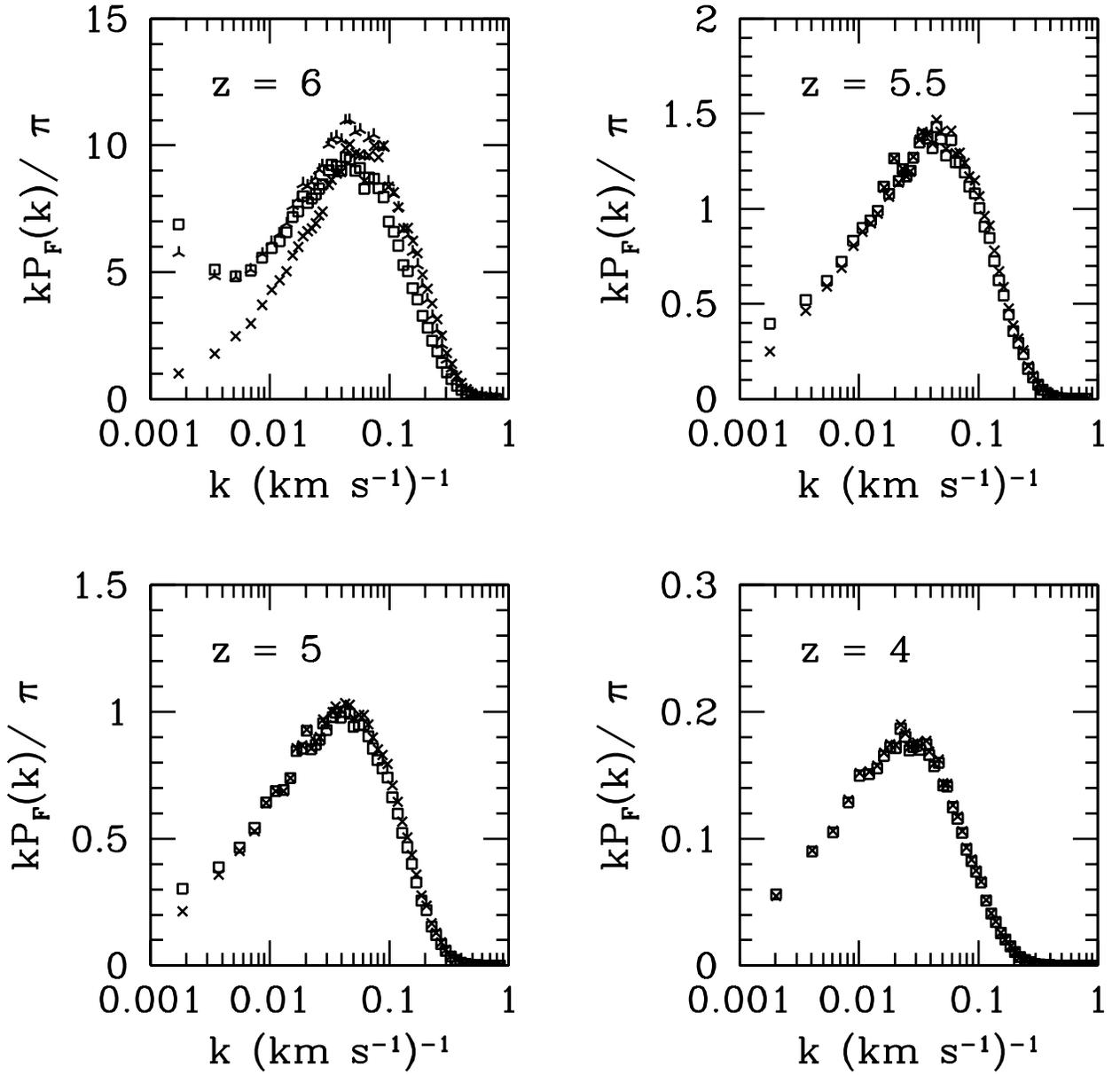}
\caption{The flux power spectrum $P_F(k)$, as a function of restframe
velocity wavenumber, for Model 4, with (square) and without ($\times$)
UV background fluctuations. Also shown is $P_F(k)$ at
$z=6$ allowing for a slight suppression due to diffuse
radiation (inverted Y). The results are shown at
$z=4$, 5, 5.5 and 6.}
\label{fig:PFk4_z46}
\end{figure*}

\begin{figure*}
\epsfbox{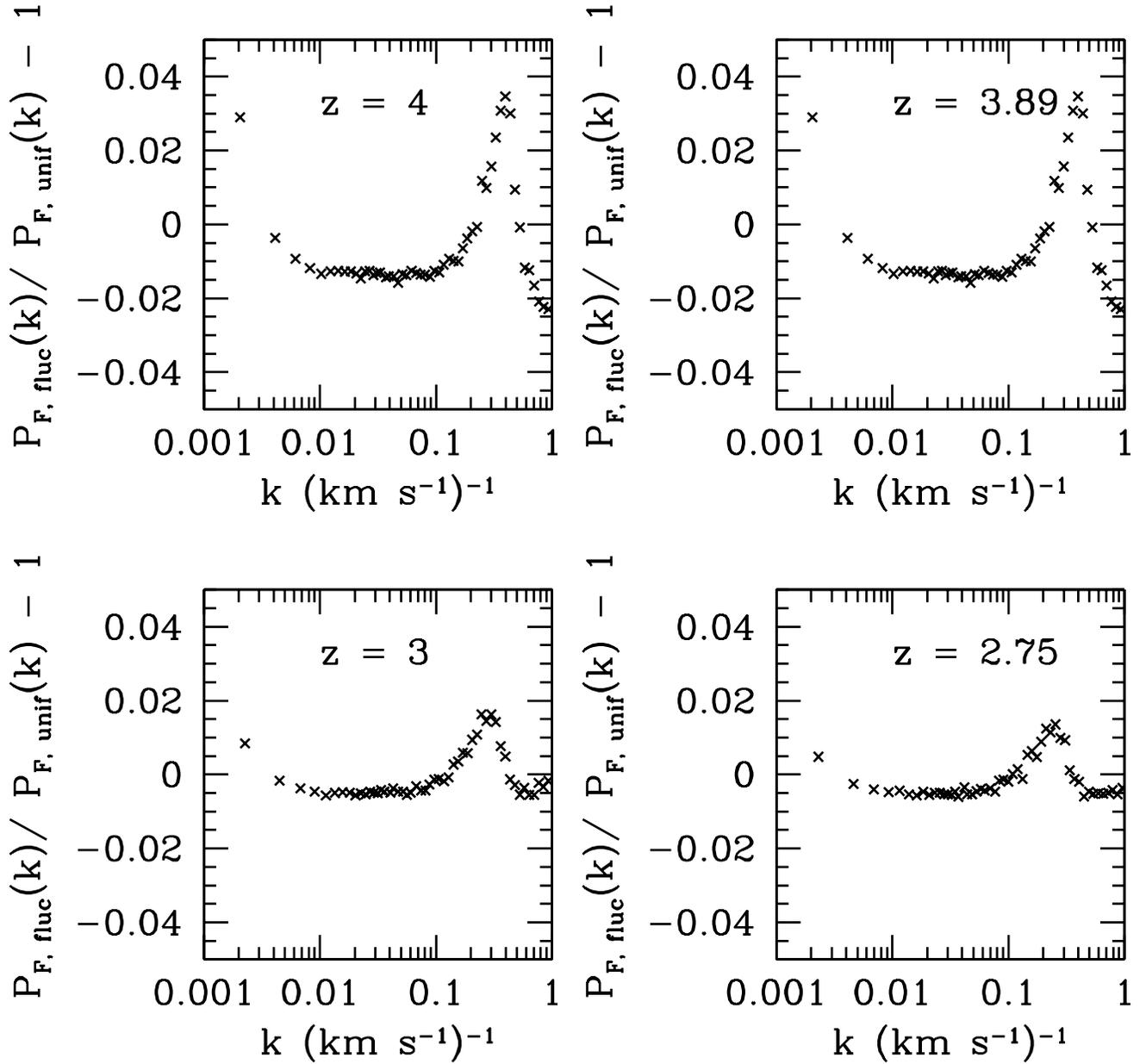}
\caption{The ratio of the flux power spectrum $P_F(k)$ with UV background
fluctuations to the value without, as a function of restframe
velocity wavenumber, for Model 4. The results are shown at
$z=2.75$, 3, 3.89 and 4.}
\label{fig:PFk4_z24}
\end{figure*}

\begin{table}
\begin{center}
\begin{tabular}{|c|c|c|} \hline
Model & $\chi^2$ & $P(>\chi^2)$ \\
\hline 
\hline
 1  & 55.1   &  0.007  \\ \hline
 2  & 38.7   &  0.192  \\ \hline
 3  & 42.3   &  0.106  \\ \hline
 4  & 40.7   &  0.139  \\ \hline
 C  & 44.9   &  0.065  \\ \hline
\end{tabular}
\end{center}
\caption{Statistical comparison betwen the model predictions and
measurements for the flux power spectrum, combining data at
$z=2.41$, 2.72, 3.00 and 3.89. No adjustment to the equation of state
or mean \Lya flux have been made.
}
\label{tab:PFk_data_comp}
\end{table}

We compute the flux power spectrum at wavenumber $k$
for the spectra according to
\begin{equation}
P_F(k) = \langle |\widehat{\delta f}|^2\rangle/ \langle f\rangle^2,
\label{eq:PFk}
\end{equation}
where $\widehat{\delta f}$ denotes the Fourier transform of
$\delta f = f-\langle f\rangle$.
In order to assess the accuracy of our predicted flux power spectra,
we compare the results for Models 1 and 4 and
the concordance model with the measurements of McDonald \etal
(2000) at $z=3.89$, 3.00 and 2.41, using spectra normalised to the identical
mean \Lya flux they report. The results are shown in Fig.~\ref{fig:PFk_data}.
The same models are compared with the data
of Croft \etal (2002b) at $2.72$ (using the simulation results at $z=2.75$).
Since Croft \etal report no mean \Lya flux
measurement for their spectra, we adopt the value in
Table~\ref{tab:aout}. We note that according to eq.(26) of
Croft \etal, Model 1 is ruled out at the $3.9\sigma$ level, while Model 4
and the concordance models at only the $1.0\sigma$ and $1.9\sigma$ levels,
respectively. At $z=2.7$, we find Model 1 fits about
as well as Model 4 and the concordance model. For $k<0.015$\skm, the
$\chi^2$ for Models 1, 4 and the concordance model, compared with the data
of Croft \etal, are, respectively, 13.8, 16.5 and 18.3. The respective
probabilities $P(>\chi^2)$ for the 8 $k-$values are 0.087, 0.036 and 0.019,
so that Model 1 is actually preferred.

In order to understand the differences between our results and those of
Croft \etal, we have performed a resolution sequence of comparisons
between the concordance model simulations and the measurements, for a fixed
equation of state (ie, leaving $T_0$ and $\gamma$ unchanged). In
Fig.~\ref{fig:PFk_data_chi2} we show how the reduced $\chi^2$ varies
with resolution, for both $k<0.015$\skm and $k<0.032$\skm.  (The
simulation values are linearly interpolated to the $k-$values of the
measured $P_F(k)$ points.) We do the comparison for mesh-scale
smoothing, Gaussian Jeans smoothing and linear Jeans smoothing. A
clear trend of diminishing $\chi^2$ with increasing resolution is
apparent in all cases, showing that the lower resolution simulations
have not yet matched the precision of the measured $P_F(k)$
values. Only at the highest resolution do the simulations begin to
converge, giving values of sufficient precision to make a meaningful
comparison with the measured values. The decrease in $\chi^2$ at
higher resolution is noteworthy, as it suggests a given model is a better
fit to the data than would be believed on the basis of lower resolution
simulations. The larger $\chi^2$ values
found for the $60\hmpc$ box for $k<0.032$\skm suggest that the
simulations have still not quite converged on these scales, at least
at a resolution of $\Delta r=58.6\, {\rm h^{-1}\, kpc}$, particularly
for $z<3$.

The results in Fig.~\ref{fig:PFk_data_chi2} also show that the
goodness-of-fit is highly sensitive to the smoothing. Adopting
Gaussian Jeans smoothing instead of smoothing on the mesh scale
results in $\chi^2=36.6$ for $k<0.015$\skm, and
$P(>\chi^2)=1.4\times10^{-5}$, in considerably poorer agreement than
the mesh-scale smoothing. The agreement is also much poorer than
predicted by Croft \etal, although we have not attempted to adjust the
mean flux or equation of state to improve the match. Similarly poorer
results are found for linear Jeans smoothing. We also compute the effect
of a change in the equation of state for the mesh-scale smoothing
case, adopting the values for $T_0$ and $\gamma$ predicted by the late
\HeII reionization model of Schaye \etal (2000). The agreement considerably
worsens, as shown in Fig.~\ref{fig:PFk_data_chi2}, resulting in values
comparable to the Jeans smoothing cases.

Without direct comparisons to full hydrodynamical simulations of
comparably high resolution, it is unclear how much artificial
smoothing should be applied to the PM results, or, indeed, how
accurately any smoothing scheme may mimic the true underlying
hydrodynamical processes. A certain amount of smoothing will always be
present in the spectra resulting from thermal broadening and peculiar
velocities in the gas, which the PM simulations are able to account
for. Indeed, there is a certain amount of degeneracy between the
Jeans-scale broadening and changes in the equation of state. The
crucial question is whether the pressure forces missing in the PM
simulations result in significantly less smoothing than should be
physically present to obtain accurate predictions for $P_F(k)$. We are
unable to address that question here, but we have demonstrated that the
successful application of PM (or HPM) simulations
for discriminating between cosmological models on the basis of
the flux power spectrum relies critically on the adopted degree
of artificial smoothing.

We compare our models with the data of McDonald \etal and Croft \etal
(2002b) in Table~\ref{tab:PFk_data_comp} combining the results at $z=2.41$,
2.72, 3.00 and 3.89 for $k<0.015$\skm. The total number of degrees of
freedom is 32. We have conservatively adopted mesh-scale smoothing
throughout. Within the context of mesh-scale smoothing, the results we
provide are upper bounds to the $\chi^2$ for each model, since we have
made no effort to improve the fits by making the several adjustments
available. Specifically, we have not:\ 1.\ adjusted $\bar\tau_\alpha$
to improve the fits, adopting instead the values given by McDonald
\etal, or from Table~\ref{tab:aout} for $z=2.7$, 2.\ adjusted the
equation of state parameters $T_0$ and $\gamma$, 3.\ accounted for the
spread in the model predictions of $P_F(k)$, or 4.\ accounted for mode-mode
correlations in the power either for the simulations or the data. We
confine the comparison to $k<0.015$\skm, since for this range the model
predictions are more stable to changes in the box size, and the range
avoids an anomalous low excursion in the measured $P_F(k)$ at $z=3$
for $k=0.0179$\skm with very small error bar.

The effect of UV background fluctuations on the flux power spectrum is shown
in Figs.~\ref{fig:PFk4_z46} and \ref{fig:PFk4_z24}. Since the UV background
fluctuations tend to somewhat suppress the mean flux (boost the mean optical
depth) (Paper I), we have re-normalised the uniform background case to the
identical mean flux found for the fluctuating background case at each redshift
to make a fair comparison. The effect of the fluctuations are most pronounced
on large scales and small scales. The small scale peak is on the scale of the
absorption line widths and is due in part to a resetting of the mean ionization
rate necessary to reproduce the same mean \Lya flux.
At intermediate scales the power tends to be somewhat suppressed. The trend
at low $k$ of decreasing power with increasing $k$ continues even
to $z=2.75$, although the effect is much reduced and
essentially vanishes by $z=2.41$. The rise at the lowest $k$
value appears to be real, not an artefact of the finite box size, based on
convergence tests applied to the concordance model (see Appendix A).
At $z=6$, we also show the small dilution effect
of introducing an additional diffuse radiation background of 50\% of the
average direct contribution from the QSO sources.

\begin{figure}
\begin{center}
\leavevmode \epsfxsize=3.3in \epsfbox{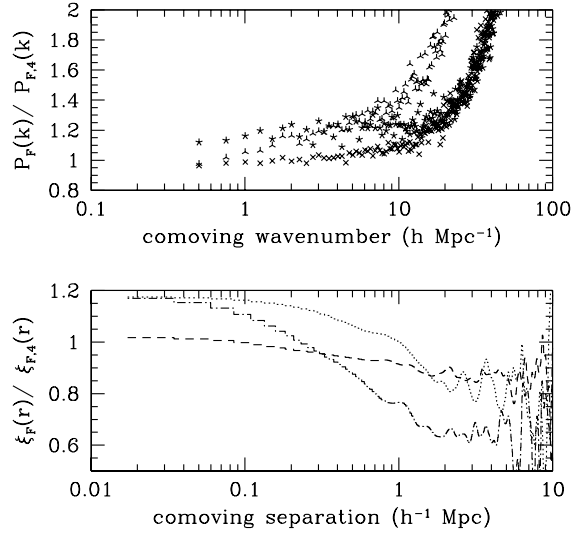}
\end{center}
\caption{(Upper panel)\ The ratio of the flux power spectrum for
Model 1 (5-pt star), Model 2 ($\times$) and Model 3 (inverted Y) to that
of Model 4 at $z=5$. (Lower panel)\ The ratio of the flux auto-correlation
function for Model 1 (dotted), Model 2 (short dashed) and Model 3
(dot-short dashed) to that of Model 4 at $z=5$.
}
\label{fig:xiFPFk}
\end{figure}

The ratios of the flux power spectra for
Models 1--3 to Model 4 are shown in Fig.~\ref{fig:xiFPFk} at $z=5$.
The differences between the models are smaller than the effect induced by the
UV background fluctuations on the largest scales (smallest wavenumbers).
The effect of the fluctuations should be discernable in $z>5$ spectra of
sufficient signal-to-noise ratio within the current precision of the measured
cosmological parameters.

\subsection{Pixel flux auto-correlation function}

\begin{figure}
\begin{center}
\leavevmode \epsfxsize=3.3in \epsfbox{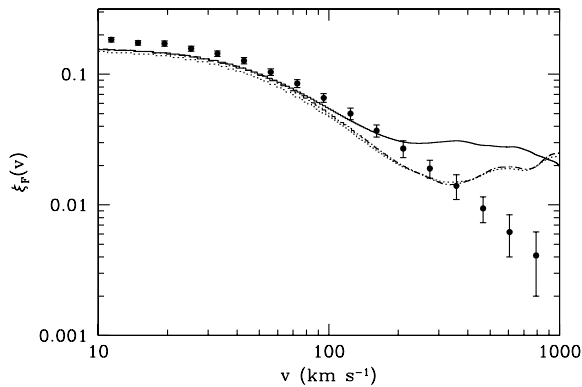}
\end{center}
\caption{The flux auto-correlation function $\xi_F(v)$ at $z=2.7$, as a
function of restframe velocity separation, for Model 4 (solid) and the
concordance model (dashed), assuming Gaussian smoothing on the mesh scale.
Also shown is the
concordance model using linear Jeans smoothing (dotted). The models are
compared with the measurements from Croft \etal (2002b).
}
\label{fig:xiF_data}
\end{figure}

\begin{figure}
\begin{center}
\leavevmode \epsfxsize=3.3in \epsfbox{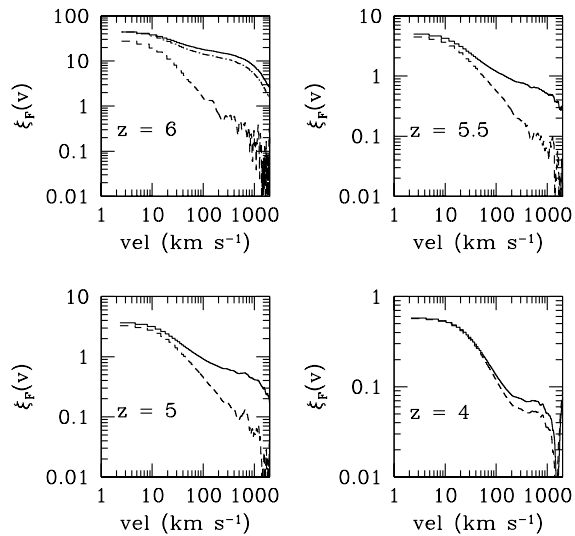}
\end{center}
\caption{The flux auto-correlation function $\xi_F$(v), as a function of rest
frame velocity separation, for Model 4,
with (solid) and without (dashed) UV background fluctuations.
Also shown is $\xi_F(v)$ at $z=6$ including UV background fluctuations,
but allowing for a slight suppression due to additional
diffuse radiation (dot-dashed line). The results are shown at
$z=4$, 5, 5.5 and 6.}
\label{fig:correl4}
\end{figure}

We compute the flux auto-correlation function at a velocity separation $v$
for the spectra according to
\begin{equation}
\xi_F(v) = \langle f(v')f(v'-v)\rangle/ \langle f\rangle^2 - 1,
\label{eq:xiF}
\end{equation}
where $f(v)$ denotes the pixel flux at velocity $v$, and
$\langle f\rangle$ is the measured mean flux at the relevant redshift.
The results are shown for Model 4 and the concordance model in
Fig.~\ref{fig:xiF_data} at $z=2.75$. The data points are from
Croft \etal (2002b) at $\langle z\rangle=2.72$. While the agreement is
close at small separations, the model estimates disagree significantly
with the measured values. The agreement worsens at large separations, likely
due to lack of numerical convergence in the simulation results. We have not
attempted to adjust the mean \Lya flux or the equation of state to improve
the match.

We compare the flux auto-correlation function with and without UV background
fluctuations in Fig.~\ref{fig:correl4}. The largest effect is at $z=6$ and
decreases with decreasing redshift. The effect almost disappears by $z=4$
for velocity separations smaller than 200\kms. We note that the convergence of
the auto-correlation function is not guaranteed for velocity separations
exceeding $\sim100$\kms. At $z=6$, we also show the small dilution effect
of introducing an additional diffuse radiation background of 50\% of the
average direct contribution from the QSO sources.

The ratios of the flux auto-correlation function for
Models 1--3 to Model 4 are shown in Fig.~\ref{fig:xiFPFk} at $z=5$.
The differences between the models are smaller than the effect induced by the
UV background fluctuations for small to moderate scales ($<100$\kms).

\section{Conclusions}

We have examined the effect of UV background fluctuations expected
from QSO sources on the power spectrum and auto-correlation function
of the \Lya forest. To do so, we first performed several sequences of
tests to assess the simulation parameters required to reach
convergence. We investigate the convergence for different amounts of
smoothing meant to mimic hydrodynamical effects, almost all for a
fixed equation of state. We note that varying the equation of state
will affect the predicted flux power spectrum, and should be included
when attempting to constrain cosmological models by comparison with
the data. We have not performed an exhaustive search in smoothing
algorithms and equations of state to achieve a best match to the
data. Our emphasis in this paper is instead to explore the impact
these have on the {\it relative} effect UV background fluctuations are
expected to have on the predicted flux power spectrum and
auto-correlation function. Our conclusions concerning these tests are
(where all length scales are comoving and velocity scales proper):

1. To reach 10\% convergence in the required \HI photoionization
rate $\Gamma_{-12}$, a (comoving) spatial resolution of
$60\hkpc$ is required.

2. We were not able to reach convergence in the 1D dark matter power spectrum
to better than 50\% on any scale. Box sizes exceeding $60\hmpc$ and/or a
spatial resolution of better than $30\hkpc$ appear to be required at $z=5$.
The convergence requirements are even more severe at lower redshifts.
Fortunately good convergence in the 1D matter power spectrum is not required
for good convergence in the predicted flux power spectrum because thermal and
Doppler broadening of the gas filters out the contribution of the high spatial
frequency modes to the \HI optical depth.

3. We show that the predicted flux power spectrum $P_F(k)$ is
senstitive to the degree of smoothing. A crucial question is how much
(or even whether) any additional smoothing to the gas variables should
be applied beyond the inherent smoothing of the spectra due to Doppler
and peculiar velocity broadening. For the minimal smoothing on the
scale of a mesh cell, the minimum resolution and box size required to
achieve 10\% convergence on (proper) scales $k<0.015$\skm is $30\hkpc$
in a $25\hmpc$ box. We were not able to achieve convergence to 10\% at
larger wavenumbers. Allowing for additional smoothing improves the
convergence.  With the application of Gaussian Jeans smoothing, it is
possible to achieve convergence on all scales $k<0.1$\skm to 10\% at
$z=5$ for a spatial resolution of at least $60\hkpc$ in a $30\hmpc$
box. With the application of linear Jeans smoothing, we are able to
achieve this level of accuracy only for $k\lsim0.01$\skm. The
convergence requirements become more severe at lower
redshifts.

4. We find that introducing Jeans Gaussian or Jeans linear smoothing
significantly flattens the predicted flux power spectrum near its
peak. Without full hydrodynamical simulations at comparable resolution
to our highest resolution simulations in a box size of at least
$25\hmpc$, it is not possible to assess whether or not the additional
Jeans smoothing, while it improves the convergence properties, also
improves the accuracy of the flux power spectrum as predicted from
pure gravity simulations.

5. We were unable to obtain convergence in the flux auto-correlation
function $\xi_F(k)$ to better than 10\% on spatial separations
exceeding $\sim3\%$ of the box size. Comparisons with the data for
this statistic appear problematic.

6. A criterion of the level of simulation convergence needed to make a
meaningful comparison with observations is the convergence in $\chi^2$
between the simulation predictions and the measured values. We find
that $\chi^2$ for the comparison between the predicted and measured
$P_F(k)$ continues to decrease with increasing resolution. For our
minimal mesh-scale smoothing case, we find differences in the reduced
$\chi^2$ at $z\approx3-4$ of $\sim50$\% on going from a resolution of
$\sim60\hkpc$ to
$\sim30\hkpc$ in a $30\hmpc$ box, and differences of $\sim20$\% on
increasing the box size to $60\hmpc$ at a fixed spatial resolution of
$\sim60\hkpc$.  Allowing for Jeans smoothing does not greatly improve
the convergence in the reduced $\chi^2$ confined to wavenumbers
$k<0.015$\skm. In particular, we find that while the concordance model
would be very strongly rejected on the basis of moderate resolution
simulations using the minimal amount of mesh-scale smoothing, the
model is rejected at only the $\lsim2\sigma$ level in our highest
resolution simulations. By contrast, when Jeans smoothing is applied,
the model is very strongly rejected at all resolutions. We have not
explored variations in the equation of state that could potentially
improve (or degrade) the agreement.

7. We find significant degeneracy between the smoothing and the
equation of state of the baryons. It is unclear how the degeneracy may
be removed without performing full hydrodynamic simulations to infer
the equation of state of the IGM by comparing with the measured small
scale fluctuations of the \Lya forest on the scale of the absorption
features. A possibility is to use the measured Doppler parameters to
constrain the equation of state (Schaye \etal 2000), but without high
resolution simulations in box sizes of at least $\sim25\hmpc$, it is
unclear whether or not the parameters deduced in this way are the
optimal ones to use for predicting the flux power spectrum. One
complication is the evident need for additional broadening at $z<3.5$
beyond simple photoionization heating (Theuns \etal 1999; Bryan \&
Machacek 2000; Meiksin, Bryan \& Machacek 2001). The discrepancy in
the predicted and measured line widths suggests the underlying physics
is still not fully understood, particularly for the moderate to low
optical depth lines (Meiksin, Bryan \& Machacek 2001).

The simulations are normalised on the basis of the average \Lya flux
transmitted through the IGM. We find that there is significantly more
accord in the measurements of this parameter
than previously recognized once existing constraints
are interpreted on a consistent statistical basis.

We assess the possibility of using measurements of the flux power
spectrum for inferring the contribution of QSO sources to the
metagalactic UV ionizing background. Because of the sparsity of QSOs
relative to alternative sources, like galaxies, the UV background
contribution from QSOs will exhibit large, spatially correlated
fluctuations. We model the fluctuations using Monte Carlo realisations
of randomly distributed QSO sources based on QSO counts from the 2dF
and SDSS QSO surveys. We estimate the attenuation lengths through the
IGM self-consistently from our simulations, requiring a background
ionization rate that recovers the measured mean \Lya flux. Because the
simulations are not able to recover the measured number of Lyman Limit
Systems for $z<4$, we add their contribution to the attenuation by hand. 

The resulting fluctuations increase the required photoionization rates
by 5-10\% over the uniform background case for $z\le5.5$ in order to
match the measured mean \Lya fluxes. This is a somewhat smaller effect
than was found in Paper I, where only the UV background fluctuations
were accounted for and not their spatial correlations. By $z=6$, we
find the trend reverses in the presence of the now very strong UV
background correlations:\ the required photoionization rate is reduced
by 10-20\% compared with the uniform background case. By contrast, we
find that the effect on the cumulative flux distribution increases when
the correlations in the UV background fluctuations are included. Previously
we found the fluctuations distorted the flux distributions by $\sim0.2$\%
for $z\ge5$. Allowing for the correlations increases the effect to 1--1.5\%,
rendering them somewhat easier to detect.

We find that the magnitudes of the {\it relative} effects on the 2-pt
flux statistics by fluctuations in the UV background are much better
converged than the absolute predictions. The ratios of the predicted
flux power spectra with and without UV background fluctuations are
converged to better than 1\% on scales $k<0.2$\skm in our simulations,
and are in agreement for Gaussian mesh, Jeans Gaussian and Jeans
linear smoothings at this level over these scales. The ratio of the
predicted auto-correlation function with and without UV background
fluctuations, however, is converged to better than 10\% only for
velocity separations $\lsim200$\kms.

We also find that the UV background fluctuations boost the flux power
spectrum at large scales ($k<0.01$\skm) and small scales
($k\sim0.2-0.4$\skm, corresponding to the widths of individual
absorption features), and suppress the power at intermediate
scales. For $z\le4$, the effects are small, at the few percent
level. At higher redshifts, however, the large scale effect grows due
to the increasing strength of the UV background correlations,
resulting in a boost in power of over 50\% by $z>5$. By $z=6$, the
background fluctuations produce an upturn in $kP_F(k)/\pi$ towards low
$k$, providing distinctive evidence for the presence of UV background
fluctuations, as such an upturn would be unphysical from large-scale
structure effects alone. Measuring such an effect at the current level
of flux detection is problematic because of the very high \Lya optical
depths at these redshifts, but it may become feasible with future
Extremely Large Telescope technology, or possibly by searching for the
signal at higher orders in the Lyman series where the optical depths are
reduced.

\bigskip
\section*{acknowledgments}

M.~White was supported by the NSF and NASA. Parts of this work were
done on the IBM-SP at the National Energy Research Scientific
Computing Center.

\appendix

\section{Convergence Tests} \label{sec:tests}

Even including only gravitational physics, simulations of the \Lya forest
are restricted by two size constraints.
High spatial resolution is required to resolve the structure of the
absorbers, while a large box size is required both to capture the
large-scale power which will affect the velocity widths of the absorption
lines as well as to obtain a fair sample of the universe.
Although a small box may be adequate to resolve the line structure, if the
box is too small the fluctuations corresponding to the scale of the box will
become nonlinear, in which case the results are no longer representative of
the cosmological model simulated.
These numerical effects have been investigated in different contexts by a
number of authors (see references in McDonald 2003 for example).

To investigate these effects of numerical resolution further we chose a
relatively large box, $30\,h^{-1}$Mpc (comoving) on a side, and simulated
the concordance cosmology using $128^3$, $256^3$ or $512^3$ particles
(corresponding to $256^3$, $512^3$ and $1024^3$ mesh cells).
At the high redshifts of interest to us the fundamental mode of the box is
safely in the linear regime, suggesting that the simulation represents a fair
sample of the universe.
To isolate the effects of numerical resolution from sample variance, we
sampled the initial conditions starting at low-$k$ and moving `outwards'
to higher $k$ so that the initial phases and amplitudes of the common modes
were identical from simulation to simulation.

\begin{figure}
\begin{center}
\leavevmode \epsfxsize=3.3in \epsfbox{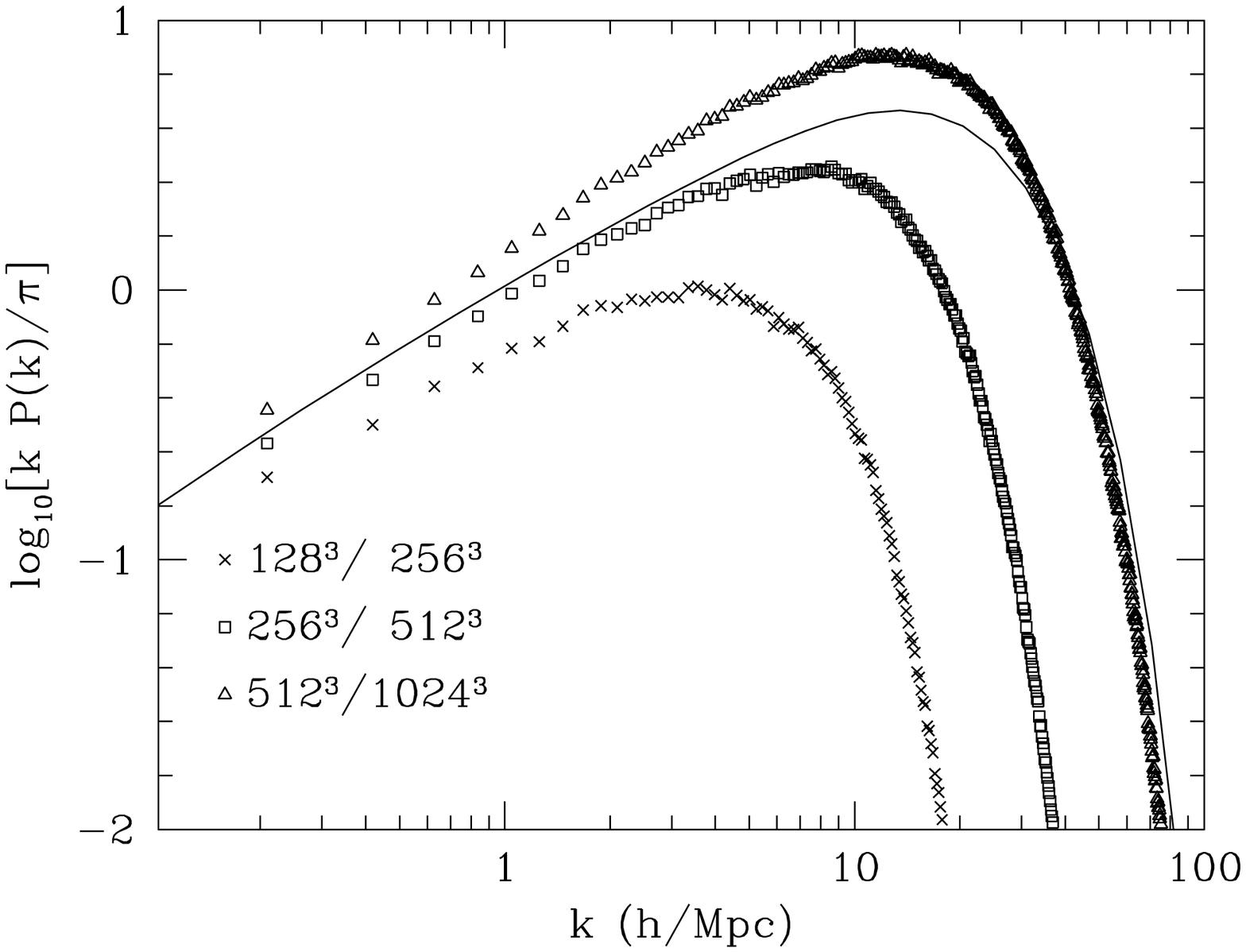}
\end{center}
\caption{The (real space) 1D matter power spectrum in the concordance
cosmology at $z=5$ from our resolution sequence with $128^3$, $256^3$
and $512^3$ particles and force meshes of $256^3$, $512^3$ and $1024^3$.
The solid line shows the predicted 1D spectrum, using the
Peacock \& Dodds (1996) fitting formulae for the 3D matter power spectrum,
Eq.~\protect\ref{eqn:pk1d}, and a Gaussian filtering of 3 mesh cells (FWHM).}
\label{fig:dk_seq}
\end{figure}

\begin{figure}
\begin{center}
\leavevmode \epsfxsize=3.3in \epsfbox{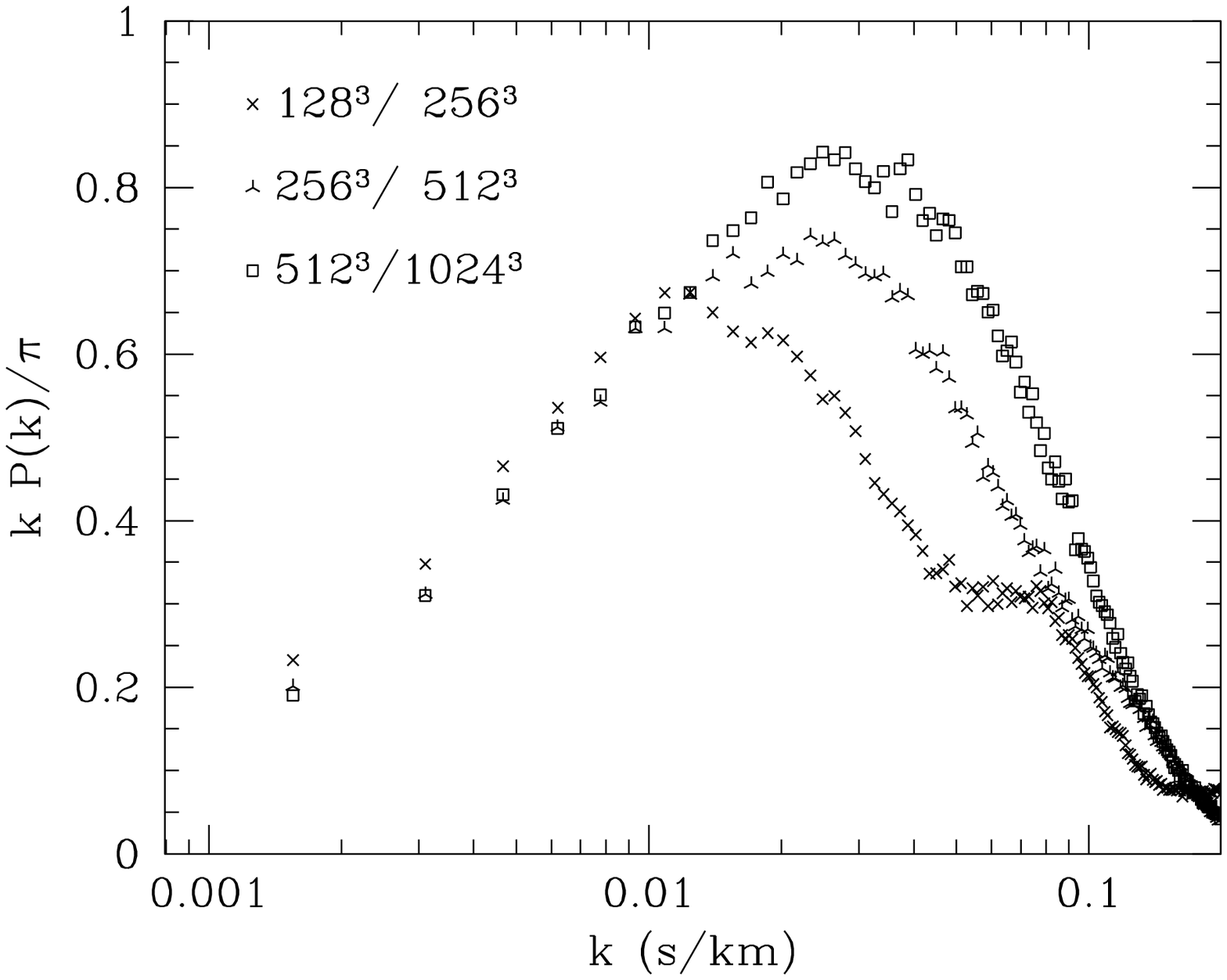}
\end{center}
\caption{The 1D flux power spectrum in the concordance cosmology at $z=5$ from
our resolution sequence with $128^3$, $256^3$ and $512^3$ particles and force
meshes of $256^3$, $512^3$ and $1024^3$.  The box size in each run is
$30\,h^{-1}$Mpc.  These results indicate that our highest resolution runs are
converged at better than the 10\% level (in power) on scales $k<0.015$\skm.}
\label{fig:pk_seq}
\end{figure}

\begin{figure*}
\vbox to 220mm{\vfil
\leavevmode \epsfxsize=3.3in \epsfbox{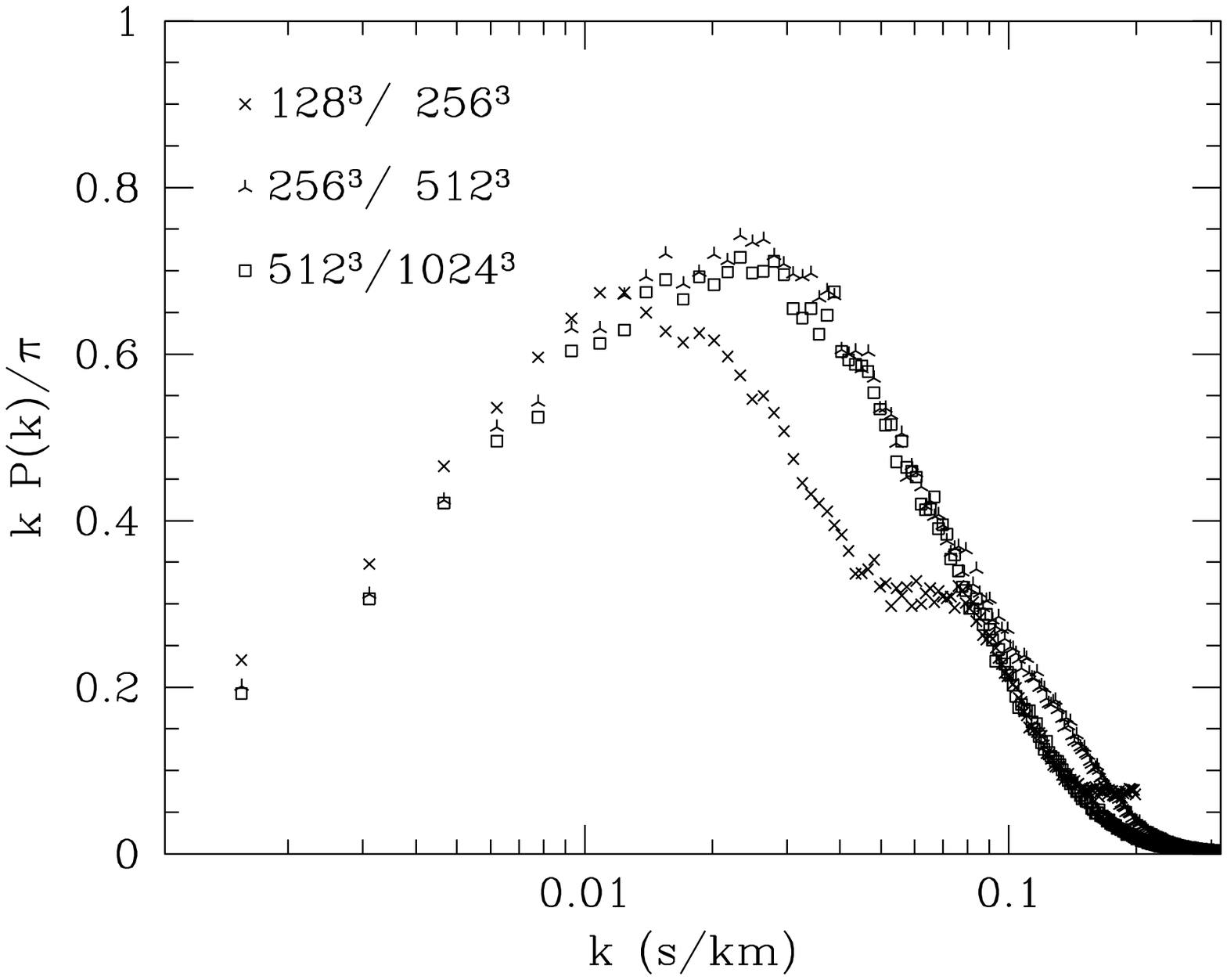}
\leavevmode \epsfxsize=3.3in \epsfbox{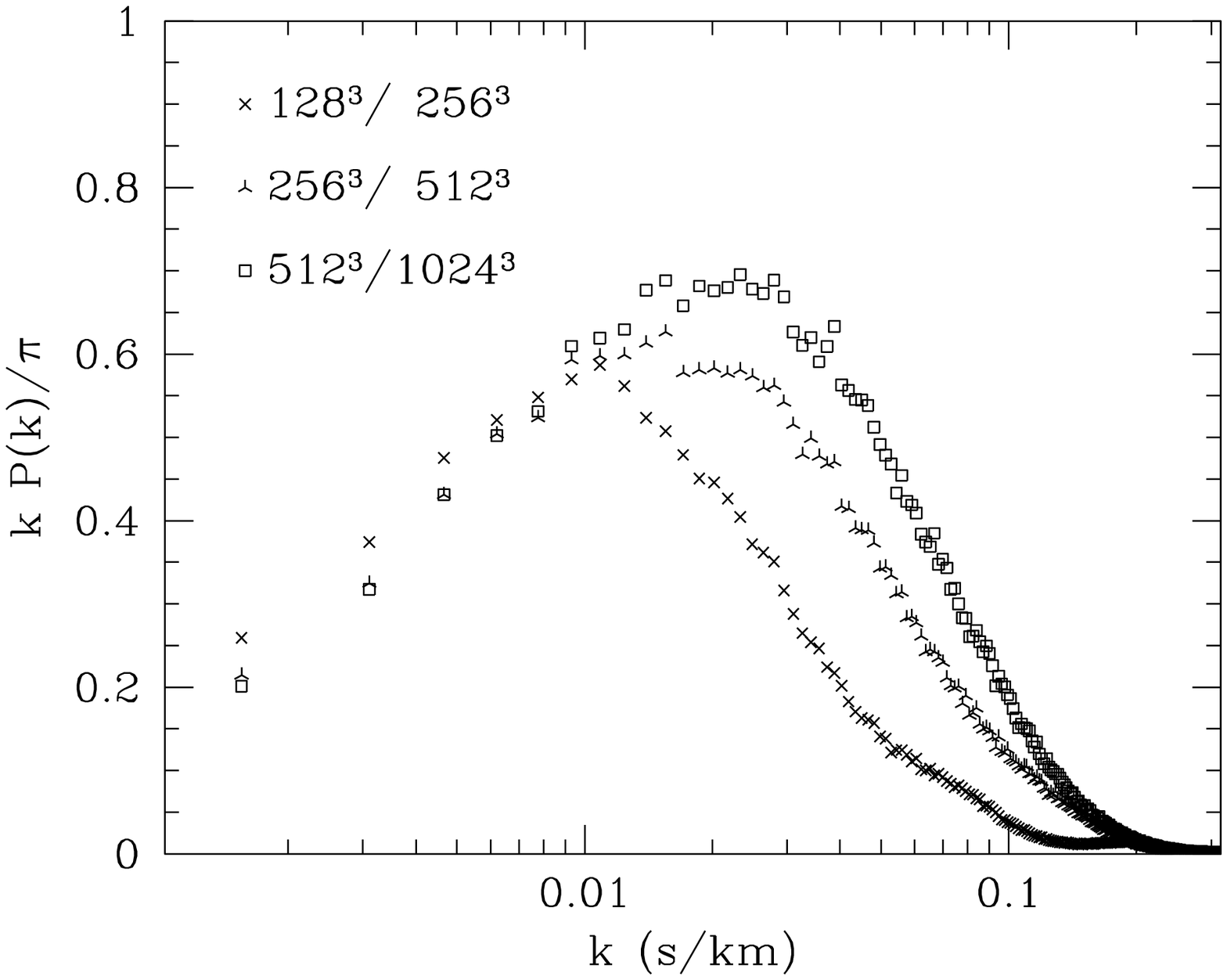}
\caption{The 1D flux power spectrum in the concordance cosmology at $z=5$ from
a resolution sequence with $128^3$, $256^3$ and $512^3$ particles and force
meshes of $256^3$, $512^3$ and $1024^3$.  The box size in each run is
$30\,h^{-1}$Mpc. (Left) The density and velocity fields are Gaussian
smoothed by the Jeans scale. (Right) The density and velocity fields are
smoothed by $(1+[k/k_J]^2)^{-1}$.}
\vfil}
\label{fig:pk_filt}
\end{figure*}

\begin{figure}
\begin{center}
\leavevmode \epsfxsize=3.3in \epsfbox{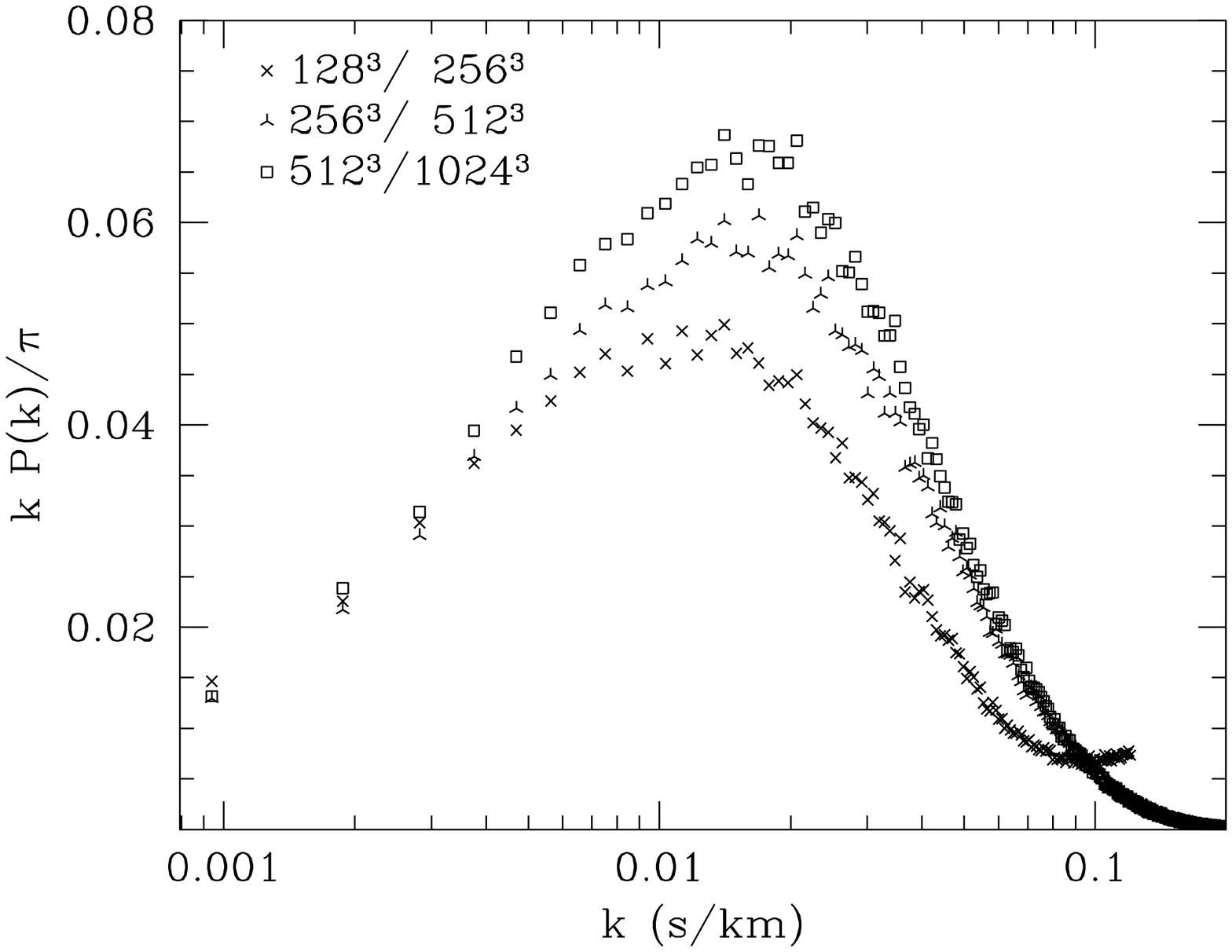}
\end{center}
\caption{The 1D flux power spectrum in the concordance cosmology at $z=3$ from
a resolution sequence with $128^3$, $256^3$ and $512^3$ particles and force
meshes of $256^3$, $512^3$ and $1024^3$.  The box size in each run is
$60\,h^{-1}$Mpc.}
\label{fig:pk_seq2}
\end{figure}

The 3D matter power spectrum (not shown) exhibits the expected behaviour,
with the higher resolution simulations having progressively more small-scale
power.  At $z=5$ the fundamental mode is safely in the linear regime, with
$\Delta^2(k_{\rm fund})\simeq 0.04$, and the mesh scale is non-linear
with $\Delta^2(k_{\rm mesh})>10$.  The fitting formula of
Peacock \& Dodds (1996) underestimates the non-linear power by
${\cal O}(50)$ per cent.
Fig.~\ref{fig:dk_seq} shows the 1D matter power spectrum, in real space.
This is an integral of the 3D spectrum over wavenumbers greater than the
1D wavenumber under consideration:
\begin{equation}
  {kP(k)\over\pi} = k \int_{k}^\infty {dy\over y^2}\ \Delta^2_{3D}(y)
  \qquad .
\label{eqn:pk1d}
\end{equation}
We find that the numerical results are relatively well fit by the
Peacock \& Dodds (1996) formalism, if we filter the 3D power spectrum,
$\Delta^2_{3D}(k)$, with a Gaussian filter of FWHM $3.2$ mesh cells
(plotted as the solid line for the highest resolution run)
to approximate the finite force resolution of the PM code.
Note that missing high-$k$ power in the lower resolution simulations biases
the 1D power spectrum low over the whole $k$-range plotted since much of
the long-wavelength power comes from aliasing of 3D short wavelength power.
We show in Fig.~\ref{fig:pk_seq} the power spectra of the flux produced by
these 3 runs using a Gaussian smoothing of one mesh cell to define the
density and velocity fields. (The flux normalisation is slightly different
than used in the text.)
{}From these results we infer that, for the relatively high redshifts of
interest, a $30\,h^{-1}$Mpc box with $512^3$ particles and a $1024^3$ force
mesh produces spectra converged to better than the $10$ per cent level.

Similar conclusions were found by McDonald (2003), though our convergence is
not quite at the level quoted therein. Part of the reason for this is the
treatment of small-scale power in the simulations. To investigate this we
modified our smoothing procedure following Gnedin \& Hui (1998)
to mimic the effects of thermal pressure on the baryonic component,
using either Gaussian or linear smoothing as described in
Section~\ref{sec:sims} above.

We show in Fig.~A3 that the Gaussian smoothing method, used
by Zaldarriaga et al.~(2001), improves the convergence. This is to be
expected since the Jeans scale here is close to the mesh size of our
intermediate resolution run. We found that pure linear theory smoothing
introduced numerical artifacts unless it was supplemented with our standard
single mesh cell Gaussian smoothing, so we show in Fig.~A3
the results of this hybrid. For the smaller runs the Gaussian smoothing
dominates, but at higher resolution the linear Jeans smoothing is suppressing
small-scale power as well.

The convergence becomes fractionally better if we increase $T_0$ or decrease
$\gamma$.  It becomes only slightly worse if we lower $T_0$ to $10^4$K.
A further sequence reducing the box size (not shown here) suggests that we can
safely reduce the volume simulated to a box of side $25\,h^{-1}$Mpc if we
are interested in the \Lya forest at high redshift. This makes the mesh scale
slightly smaller, capturing the features in the forest slightly better, thus
we have used $25\,h^{-1}$Mpc thoughout.

Finally, we show in Fig.~\ref{fig:pk_seq2}\ a convergence study at lower
redshift ($z=3$) in a larger box ($60\,h^{-1}$Mpc). The power spectrum in
the larger box follows closely that of the smaller box at half the resolution.
It thus departs significantly from that in the $30\,h^{-1}$Mpc box at
$k\simeq 0.01\,{\rm km}^{-1}\,{\rm s}$ for the highest resolution runs,
slightly earlier than might be expected from Fig.~\ref{fig:pk_seq} due to
the increased non-linear scale.

\begin{figure}
\begin{center}
\leavevmode \epsfxsize=3.3in \epsfbox{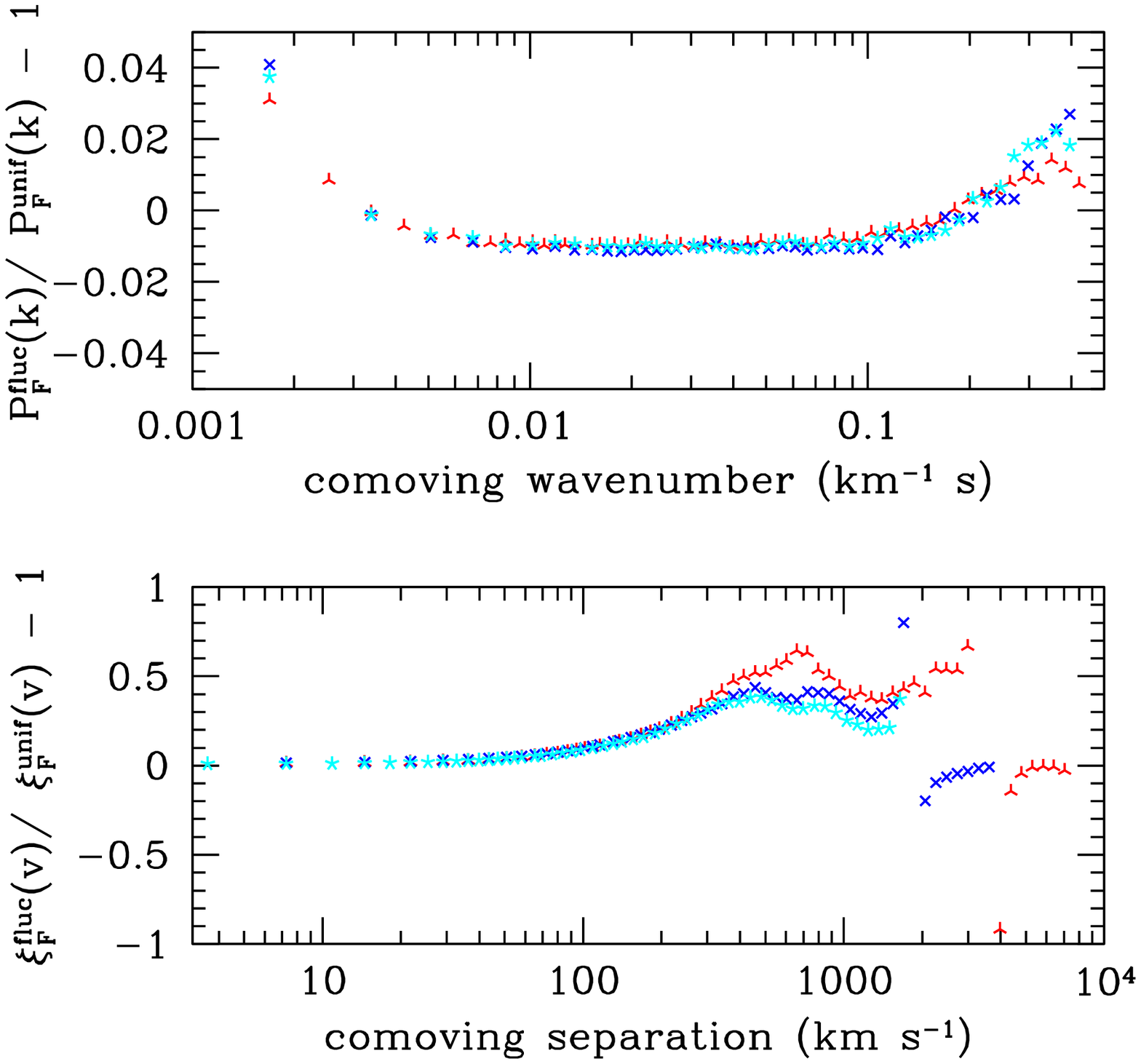}
\end{center}
\caption{The ratios of the (upper panel) 1D flux power spectrum and
(lower panel) flux auto-correlation function, with and without UV background
fluctuations, in the concordance cosmology at $z=4$ from
a resolution sequence with comoving grid size $\Delta r=58.6\hkpc$,
in boxes of comoving sides $30\hmpc$ ($\times$) and $60\hmpc$ (inverted Y), and
with a comoving grid size $\Delta r=29.3\hkpc$ in a box of
comoving side $30\hmpc$ (5-pt star).}
\label{fig:JMCpk_seq}
\end{figure}

In Fig.~\ref{fig:JMCpk_seq}, we show a convergence sequence at $z=4$
for models including fluctuations in the UV background. The UV
background fluctuations are based on the QSO counts with $\beta_1=3.2$
described in the text, with an added diffuse component increasing the
ionization rate by 40\% to reproduce the measured average flux. The
rise in $P_F(k)$ at the lowest $k$-values appears to be real. The
effect of the UV background fluctuations on the flux auto-correlation
becomes poorly determined for velocity separations exceeding
300\kms.

\begin{figure}
\begin{center}
\leavevmode \epsfxsize=3.3in \epsfbox{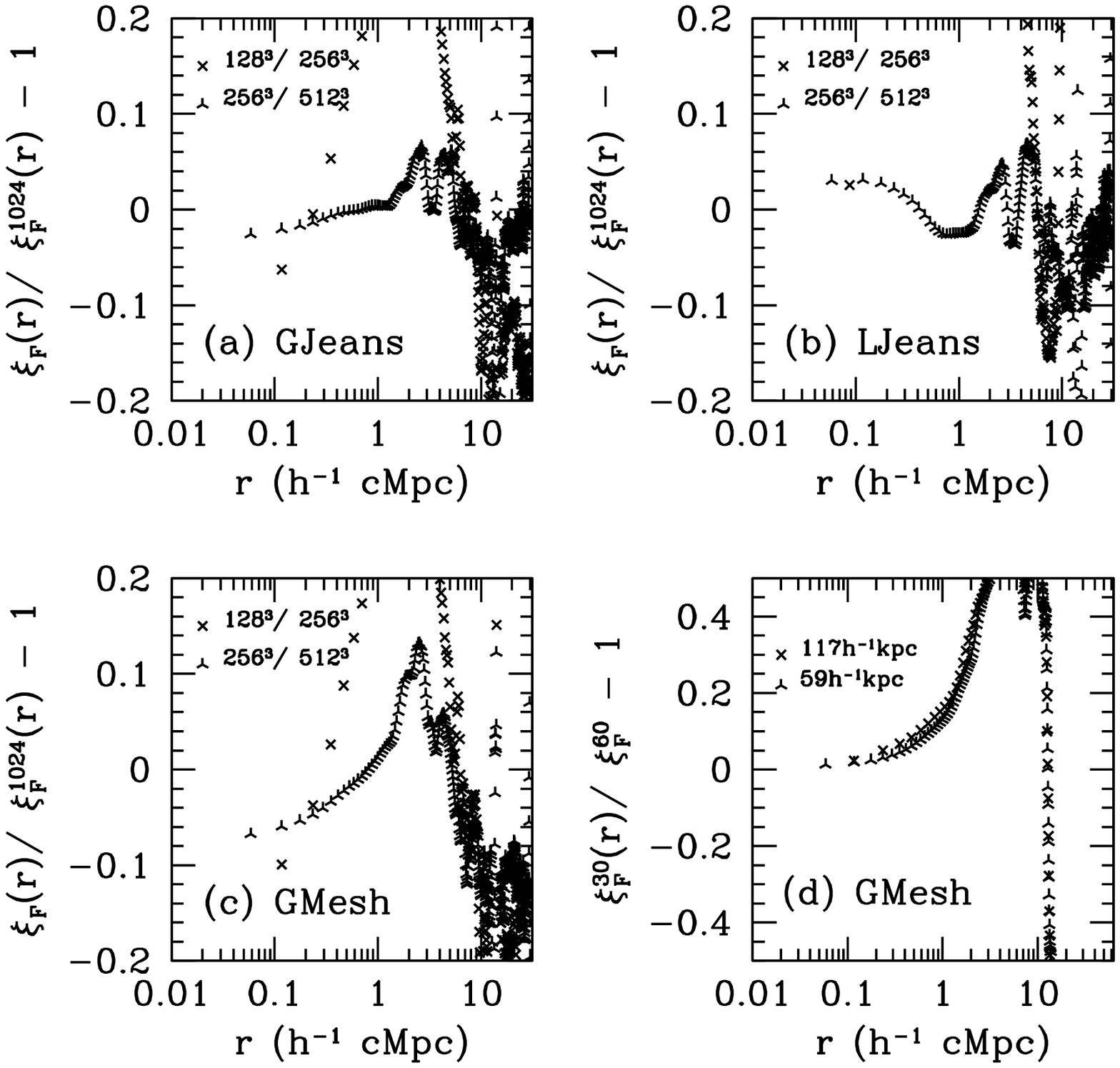}
\end{center}
\caption{The error in the 1D flux auto-correlation function in
the concordance cosmology at $z=4$ from (panels a--c) a resolution sequence
in a box of comoving size $30\,h^{-1}$Mpc with $128^3$ and $256^3$ particles
and force meshes of $256^3$ and $512^3$, compared with $512^3$
particles and a force mesh of $1024^3$, and from (panel d) a box size sequence 
comparing boxes of comoving size $30\,h^{-1}$Mpc with $60\,h^{-1}$Mpc for
(comoving) force mesh resolutions of $\Delta r = 117\, {\rm h^{-1}\, Mpc}$
and $58.6\, {\rm h^{-1}\, Mpc}$.
}
\label{fig:xiF_conv}
\end{figure}

We show in Fig.~\ref{fig:xiF_conv} that the Gaussian and linear Jeans smoothing
improve the convergence of the flux auto-correlation function over the Gaussian
single-mesh scale smoothing. For $256^3$ particles on a $512^3$ force mesh, the
Jeans smoothings give convergence to better than 5\% on comoving separations
of $r<3\,h^{-1}$Mpc, or 10\% of the box size. The single-mesh Gaussian
smoothing converges to only 15\% on these scales. Also shown is a comparison
of the single-mess Gaussian smoothing for comoving box sizes of $30\,h^{-1}$Mpc
and $60\,h^{-1}$Mpc, at two spatial resolutions,
showing convergence to 10\% only for comoving separations
of $r<1\,h^{-1}$Mpc, or 3\% of the box size.

While the smoothing schemes introduced above improve convergence, this does
not guarantee that they coverge to the correct power spectra, as would be
computed using a full hydrodynamical scheme. A direct comparison is still
prohibitively computationally expensive because of the wide range in
lengthscales required both to resolve the non-linear structures into which
the bulk of the IGM collapses and to capture sufficient large-scale power
to obtain an accurate computation of the flux power spectrum and flux
auto-correlation function. The most we can demonstrate is that the results are
not excessively sensitive to the smoothing scheme.

\begin{figure}
\begin{center}
\leavevmode \epsfxsize=3.3in \epsfbox{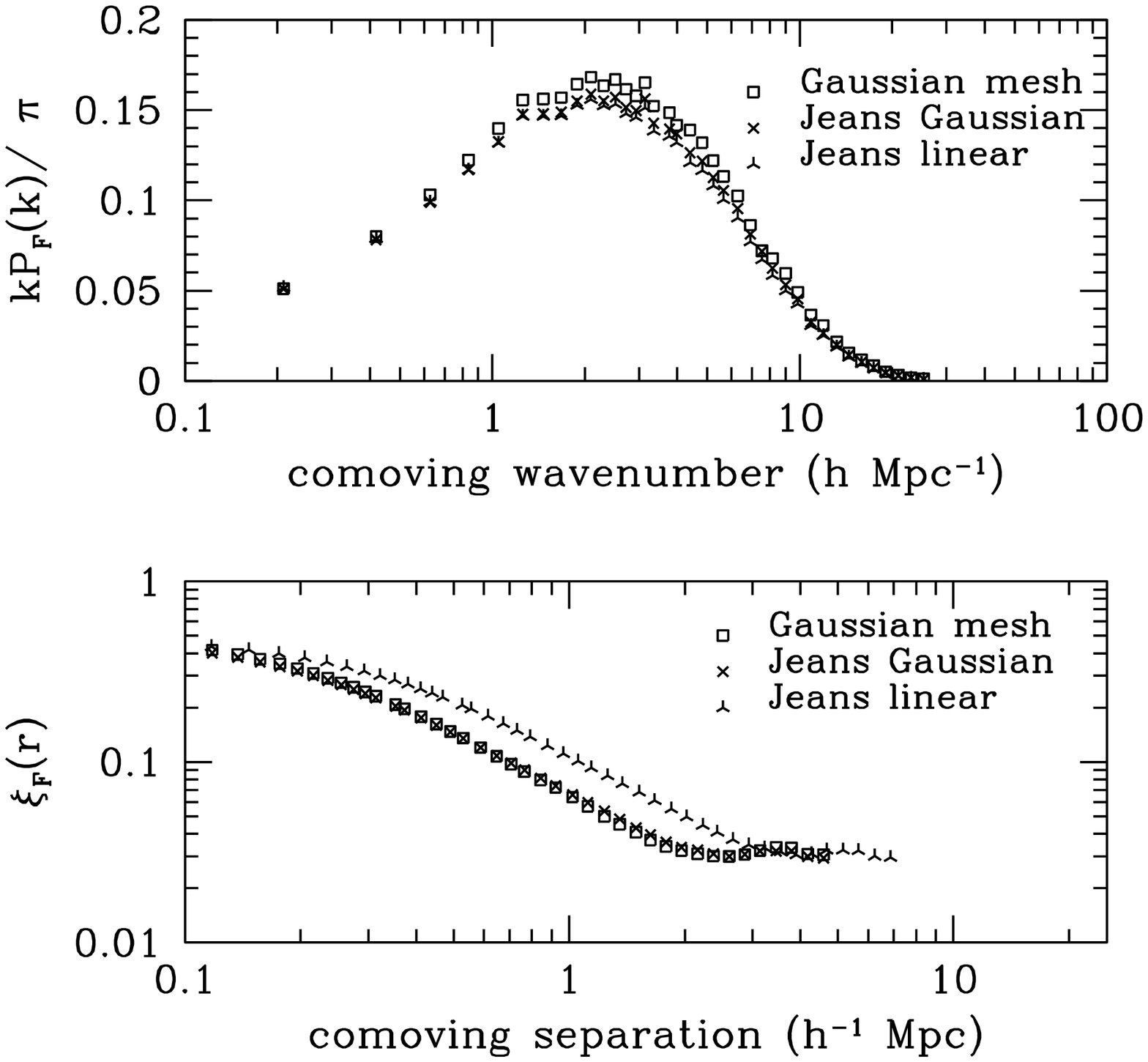}
\end{center}
\caption{The 1D flux power spectrum (upper panel) and flux auto-correlation
function (lower panel) in the concordance cosmology at $z=4$ for Gaussian
smoothing on the mesh size, Jeans Gaussian smoothing and Jeans linear
smoothing. The box size in each run is $30\,h^{-1}$Mpc.
}
\label{fig:xiF_PFk_z4}
\end{figure}

In Fig.~\ref{fig:xiF_PFk_z4}, we compare the three smoothing methods
for the concordance model at $z=4$. The power spectra for both Jeans
smoothings agree to within 1--2\%, although they are suppressed in power
compared with the Gaussian single-mesh smoothing by as much as 8\% near
the peak. By contrast, both Gaussian smoothings result in flux auto-correlation
functions agreeing to 3\%, while the linear Jeans smoothing results in an
excess of correlations of as much as 60\%. These results leave ambiguous which
smoothing method may be most accurate. For simplicity, we retain the
mesh-scale Gaussian smoothing for the main results in the text, but make
reference to alternative smoothings.

\begin{figure}
\begin{center}
\leavevmode \epsfxsize=3.3in \epsfbox{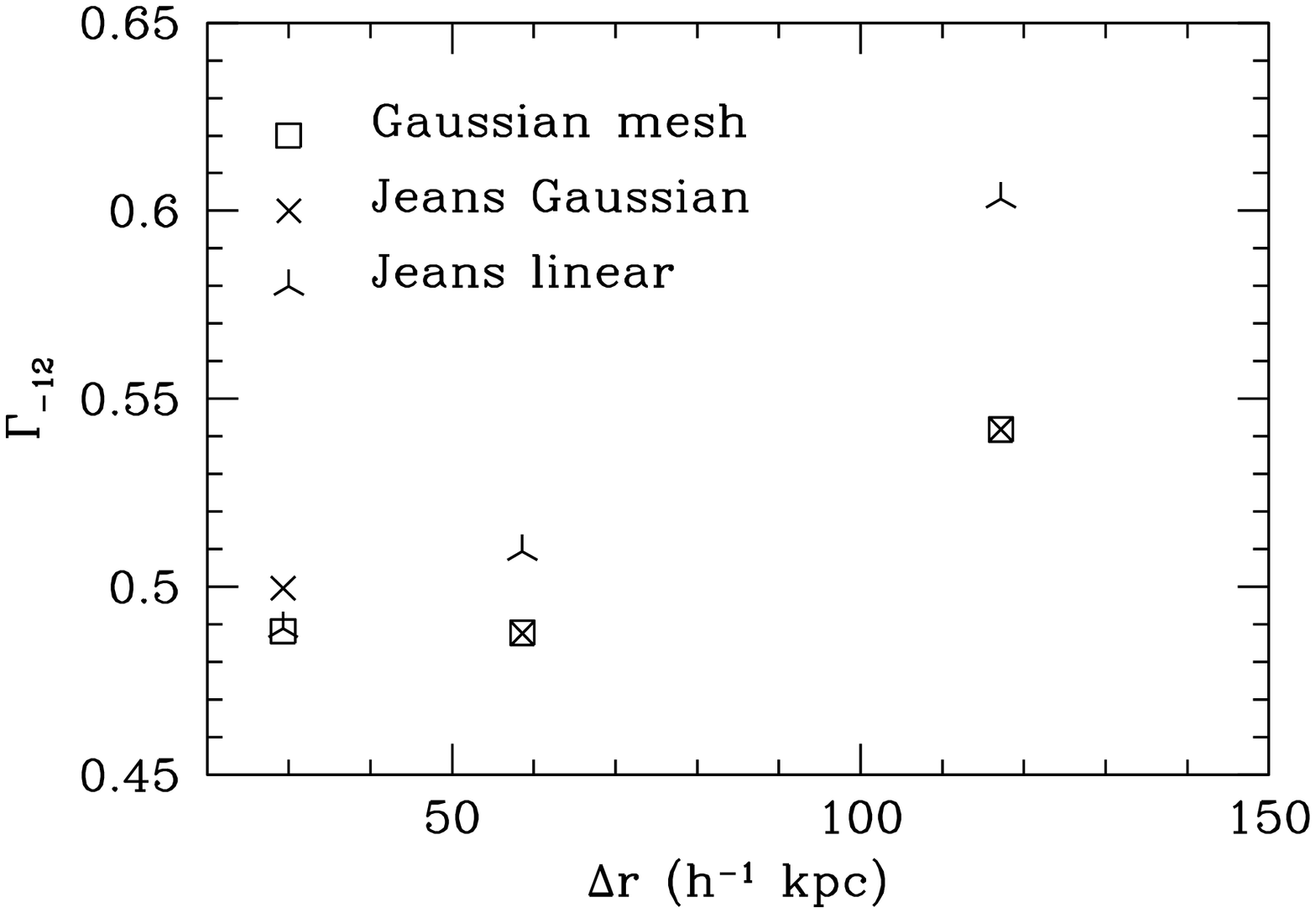}
\end{center}
\caption{The convergence of the required metagalactic ionization rate
$\Gamma_{-12}$, in units of $10^{-12}\,{\rm s}^{-1}$, at $z=4$ for force
mesh (comoving) resolutions of $\Delta r=29.3\, {\rm h^{-1}\, kpc}$,
$58.6\, {\rm h^{-1}\, kpc}$ and  $117\, {\rm h^{-1}\, kpc}$, for Gaussian
smoothing on the mesh size, Jeans Gaussian smoothing and Jeans linear
smoothing. Results shown are for the concordance model. The comoving
box size in each run is $30\,h^{-1}$Mpc.
}
\label{fig:G12_conv}
\end{figure}

The different smoothing methods result in different rates of convergence
of the metagalactic ionization rate $\Gamma_{-12}$ required to
match the measured mean \Lya flux, as shown in Fig.~\ref{fig:G12_conv}.
For a force mesh of $1024^3$, the different methods yield ionization rates
agreeing to 2\%. This is comparable to the difference between meshes of sizes
$512^3$ and $1024^3$ for each smoothing, except for the linear Jeans smoothing,
for which the difference is 4\%.

\section{Mean flux} \label{sec:meanflux}

The simulations leave unspecified the ionization level of the IGM. As
a consequence, the flux normalisation of the spectra is undetermined.
We fix the normalisation of the simulations by
adjusting $A$ in eq.~\ref{eq:Adef} to match the measured mean \Lya flux
values as given by Fan \etal (2002) in their Figure 1 for $z\ge4$,
as tabulated in Table \ref{tab:aout}. For $z<3.89$, we estimate the mean flux
from a set of published Keck HIRES spectra, as described below.

\begin{figure}
\begin{center}
\leavevmode \epsfxsize=3.3in \epsfbox{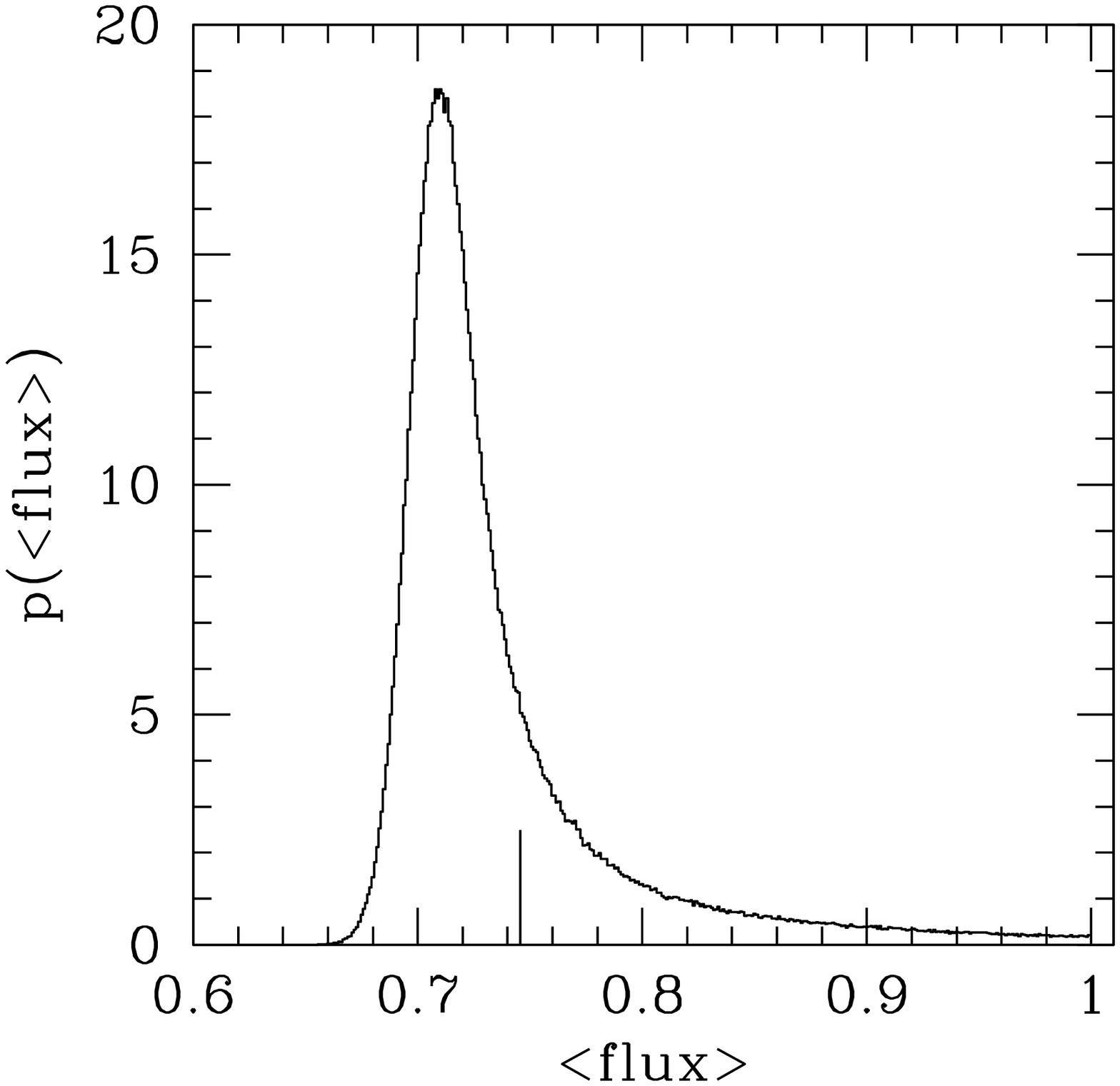}
\end{center}
\caption{The probability distribution of the mean absorbed \Lya flux
at $z=2.72$ according to the estimate of PRS. The long vertical tick mark
indicates the average mean flux.
}
\label{fig:taueff_dist_PRS93}
\end{figure}

The measurements of the mean flux have proven difficult, largely due
to uncertainties in the QSO continuum across the absorbing region.
Deriving error bars on the estimates is even more difficult because of
difficulties in measuring the error on the continuum estimate, and because
of flux correlations between pixels. PRS
used bootstrap resampling to estimate the errors on their fit to the
mean \Lya optical depth $\bar\tau_\alpha$, obtaining
$\bar\tau_\alpha=A(1+z)^{1+\gamma}$, with $\gamma=2.46\pm0.37$ and
$A=0.0175-0.0056\gamma\pm0.0002$. The interpretation of the error is
not completely clear. Seljak, McDonald \& Makarov (2003) argue that, if the
errors from PRS are interpreted as Gaussian, then the PRS fit results in
both a bias and large error estimate for $\bar\tau_\alpha$. Seljak \etal find
at $z=2.72$, $\langle\bar\tau_\alpha\rangle=0.29\pm0.14$, where $\langle\dots
\rangle$ denotes an average over Gaussian distributions for $A$ and $\gamma$,
while using the mean values of $A$ and $\gamma$ would instead give
$\bar\tau_\alpha=0.35$. The error they quote, however, is the statistical
{\it rms} on $\bar\tau_\alpha$, which carries little probabilistic meaning
since the resulting probability distribution of
$\bar\tau_\alpha$ is highly skewed. Moreover, they allowed for unphysical
values of $\bar\tau_\alpha<0$ in their averaging. Truncating the distribution
to $\bar\tau_\alpha\ge0$ yields an {\it rms} of 0.09 instead, which we estimate
from a Monte Carlo calculation. In terms of an equivalent Gaussian $1\sigma$
error spread, defined so that the probabilities of obtaining a value $1\sigma$
above or $1\sigma$ below the mean are each 0.16, the equivalent 1$\sigma$ error
spread in $\bar\tau_\alpha$ is
$\langle\bar\tau_\alpha\rangle=0.30^{+0.06}_{-0.05}$, truncating to
$\bar\tau_\alpha\ge0$.

Relating the mean optical depth to the mean absorbed flux and its
error is not straightfoward. Instead it is better to start with the
distributions for $A$ and $\gamma$ and estimate
$\langle\exp(-\bar\tau_\alpha)\rangle$ directly. The resulting
probability distribution for the mean flux at $z=2.72$ is shown in
Figure~\ref{fig:taueff_dist_PRS93}. The peak of the distribution is at
$\exp(-\bar\tau_\alpha)=0.71^{+0.03}_{-0.02}$, where the error bars
refer to the range in the mean flux enclosing 68\% of the
probability. The mean is $\langle\exp(-\bar\tau_\alpha)\rangle=0.75\pm0.07$,
where the error is the {\it rms}, and the distribution was truncated to
$\bar\tau_\alpha\ge0$. Adopting the $1\sigma$ equivalent probabilistic
interpretation of the error, as above, gives instead a mean of
$\langle\exp(-\bar\tau_\alpha)\rangle=0.75_{-0.05}^{+0.04}$. Because
of the skewed nature of the probability distribution, the estimate of
the error for any quantity depending on the mean flux must take into
account the full probability distribution of the mean flux.

\begin{table}
\begin{center}
\begin{tabular}{|c|c|c|c|c|} \hline
$z$ & PRS & SS87 & KCD & this paper \\ \hline\hline
4.0 & $0.46^{+0.06}_{-0.08}$ & $0.48\pm0.05$ & $0.49\pm0.01$ & $0.47\pm0.03$ \\
3.89 & $0.48^{+0.05}_{-0.08}$ & $0.51\pm0.05$ & $0.51\pm0.01$ & $0.48\pm0.02$ \\
3.0 & $0.69^{+0.04}_{-0.05}$ & $0.71\pm0.03$ & $0.71\pm0.01$ & $0.70\pm0.02$ \\
2.75 & $0.74^{+0.04}_{-0.05}$ & $0.76\pm0.03$ & $0.76\pm0.01$ & $0.74\pm0.04$ \\
2.41 & $0.80^{+0.03}_{-0.04}$ & $0.82\pm0.02$ & $0.82\pm0.01$ & ------- \\
\end{tabular}
\end{center}
\caption{Estimates for $\langle\exp(-\tau)\rangle$ using fits to
$\bar\tau_\alpha$ from Press, Rybicki \& Schneider (1993) (PRS), Zhang \etal
(1997) as fit to the $D_A$ values of Steidel \& Sargent (1987) (SS87), and
Kim, Cristiani \& D'Odorico (2001) (KCD). Also shown are the values adopted
in this paper.}
\label{tab:fluxa}
\end{table}

For $z<4$, we use the published Keck HIRES spectra used by Meiksin,
Bryan \& Machacek (2001) for testing models of the \Lya forest. These spectra
cover the redshift range $2.5<z<3.7$. We estimate the mean flux from
each spectrum over a narrow redshift interval ($\Delta z = 0.1-0.3$)
centred at each redshift considered, and average the results, adopting
the error in the mean as our error estimate. We also allow for small
systematic offsets in the continuum level (typically 1--4\%), as provided by
Meiksin \etal for the two best-fitting models to the flux distribution
(which are $\Lambda$CDM models similar to
those considered in this paper). We estimate the error in the systematic
offset from the spread between the best-fitting models (1--2\% of the
continuum level), and add this linearly to the error in the mean. (Since the
error is systematic rather than statistical, adding the error in quadrature
would be inappropriate.)

For $z=3.89$, we adopt the value of McDonald \etal (2000) of $0.48\pm0.02$.
For $z=4.0$, we use the best-fit of Songaila \& Cowie (2002) to the measured
redshift trend (their eq.[19]), to obtain a value for the mean flux of 0.47.
We also linearly extrapolate the values of Songaila \& Cowie (2002)
at $\langle z\rangle=4.34$ and 4.07 down to $z=4.0$, obtaining a mean flux
of $0.36\pm0.03$. The difference appears due to a single outlying measurement
which substantially lowers the mean. We shall consider both values, noting
that the larger one better matches the trend in Table~\ref{tab:fluxa}.

For comparison, we also show the estimates, using the above procedure as
applied to the fit to $\bar\tau_\alpha$ of Press \etal, based on similar
estimates from Kim, Cristiani \& D'Odorico (2001) with $A=0.0144-0.00471\gamma$
and $\gamma=2.43\pm0.17$ (which we infer from their error estimates
for $A$ and $\gamma$), and the fit by Zhang, Anninos, Norman \& Meiksin (1997)
to the flux decrement values ($D_A$) of Steidel \& Sargent (1987),
$A=0.0028\pm0.0004$, $\gamma=2.46$ (fixed). The errors provided correspond
to the equivalent $1\sigma$ Gaussian errors, as described above.

The results are tabulated in Table~\ref{tab:fluxa}.
All the estimates are found to be in accord, including those of Press \etal
This is noteworthy in that the estimates that had been made using the best
value for $\bar\tau_\alpha$ from Press \etal
generally gave values significantly higher than measured using high
resolution spectra. Estimating the mean flux including the errors as
we have done shows that these results are actually in agreement with
the estimates from the higher resolution data. It is also noteworthy
that the estimates based on the moderate resolution data of Steidel \&
Sargent are nearly identical to those using the $\bar\tau_\alpha$
fit of Kim \etal, which is based on a compilation of several high resolution
spectra. The agreement in the determinations of the mean flux between the
moderate and high resolution spectra suggests that moderate resolution spectra,
like those of the Sloan Digital Sky Survey, should, in principle, be adequate
for measuring the mean flux. At attempt to measure the mean flux from
the Sloan QSO survey was made by Bernardi \etal (2003). The values found
at $z=2.41$, 2.75, 3.0 and 3.89 are, respectively (as given in their Fig.~21),
$\langle\exp(-\tau)\rangle=0.74\pm0.03$, $0.71\pm0.01$, $0.65\pm0.01$, and
$0.36\pm0.02$. These all lie systematically low compared with the values listed
in Table~\ref{tab:fluxa}. The reasons for the discrepancy are unclear.

Since the mean flux is the quantity of interest for normalising the
\Lya forest simulations, we use direct measurements of the mean flux
to normalise our simulations rather than estimates of the mean \Lya
optical depth. Historically, $\bar\tau_\alpha$ was of interest in the
context of models for which the absorption was produced by discrete
aborption clouds rather than modulations in a continuous medium. For
the purposes of studies of the IGM under the current continuum model of
its structure, $\bar\tau_\alpha$ has lost most of its significance; the
mean flux is almost always to be preferred. An exception is in
estimating the contribution of known discrete systems to the mean
flux. Of particular interest is the contribution from Damped \Lya
Absorbers (DLAs). The contribution of discrete absorption systems
to $\bar\tau_\alpha$ is given by (eg, Meiksin \& Madau 1993; PRS)
\begin{equation}
\bar\tau_{\rm eff}=\frac{1+z}{\lambda_\alpha}\int d\NHI \frac{\partial^2
N}{\partial\NHI \partial z} w(\NHI),
\label{eq:tauDLA}
\end{equation}
where $w(\NHI)$ is the equivalent width
corresponding to a system with \HI column density $\NHI$, and
$\lambda_\alpha$ is the (rest) wavelength of the \Lya transition.
Here, $\partial^2 N/(\partial \NHI\partial z)$ is the \HI column
density distribution at redshift $z$, assumed to vary slowly over the
width of an absorption feature. The $\NHI$ distribution is estimated to
be a power law $\propto\NHI^{-\beta}$ with $\beta=1.5$, with a possible
high $\NHI$ cut-off (Storrie-Lombardi,
Irwin \& McMahon 1996). Since DLAs are on the square-root part of the curve of
growth ($w\propto\NHI^{1/2}$), $\bar\tau_{\rm DLA}$ is logarithmically
divergent at the upper end. Even if the distribution steepens at the
upper end, as suggested by Storrie-Lombardi \etal,
the statistics are so poor that the contribution of DLAs to
$\bar\tau_\alpha$ is highly uncertain. Seljak \etal claim to have
adjusted their error estimate of $\bar\tau_\alpha$ to allow
for DLAs, although they give no details of their procedure. In view
of the uncertainty in $\bar\tau_{\rm DLA}$, the most reliable approach is to
use only spectral regions that are clean of DLA contamination to
estimate the mean flux.


\end{document}